%% file: kelly_alves_2024.tex
\documentclass[preprint,12pt]{elsarticle}
\usepackage{fullpage}
\usepackage{setspace} 
\usepackage{amsmath}  
\usepackage{bm}
\usepackage{graphicx} 
\usepackage{subfigure}
\usepackage{color}
\usepackage{natbib}
\usepackage{physics}
\input{FXG_Macros}

\begin{document}
\begin{frontmatter}
\title{A Nonhydrostatic Mass-Conserving Dynamical Core for Deep Atmospheres of Variable Composition}

\author[1]{James F. Kelly \corref{cor1}}
\ead{james.f.kelly203.civ@us.navy.mil}

\author[2]{Felipe A.\ V.\ de Bragan\c{c}a Alves\corref{cor2}}
\ead{ fbalves@ups.br}

\author[1]{Stephen D. Eckermann}
\ead{stephen.eckermann.civ@us.navy.mil}

\author[2]{Francis X. Giraldo}
\ead{fxgirald@nps.edu}

\author[3]{P. Alex Reinecke}
\ead{patrick.a.reinecke.civ@us.navy.mil}

\author[1]{John T. Emmert}
\ead{john.t.emmert.civ@us.navy.mil}

\cortext[cor1]{Corresponding author}
\cortext[cor2]{Now at Instituto de Matem\'{a}tica e Estat\'{i}stica, Universidade de S\~{a}o Paulo, Brazil}

\affiliation[1]{organization={Space Science Division, U.S. Naval Research Laboratory},
            city={Washington, DC},
            country={U.S.}}
\affiliation[2]{organization={Department of Applied Mathematics, Naval Postgraduate School},
            city={Monterey},
            state={CA},
            country={U.S.}}
\affiliation[3]{organization={Marine Meteorology Division, U.S. Naval Research Laboratory},
            city={Monterey, CA},
            country={U.S.}}

\begin{abstract}
This paper develops and tests a deep-atmosphere, nonhydrostatic dynamical core (DyCore) targeted towards ground-thermosphere atmospheric prediction using the spectral element method (SEM) with Implicit-Explicit (IMEX) and Horizontally Explicit Vertically Implicit (HEVI) time-integration.  Two versions of the DyCore are presented and tested, each based on a different formulation of the specific internal energy and continuity equations, which, unlike standard potential temperature formulations, are valid for variable composition atmospheres.  The first version, which uses a product-rule (PR) forms of the continuity and specific internal energy equation, contains an additional pressure dilation term and does not conserve mass.  The second version, which does not use the product-rule (no-PR) in the continuity and specific internal energy, contains two terms to represent pressure dilation and conserves mass to machine precision regardless of time truncation error.  The pressure gradient and gravitational forces in the momentum balance equation are reformulated to reduce
numerical errors at high altitudes.  These new equation sets were implemented in two SEM-based atmospheric models: the Nonhydrostatic Unified Model of the Atmosphere (NUMA) and the Navy Environmental Prediction sysTem Using a Nonhydrostatic Engine (NEPTUNE). Numerical results using both a deep-atmosphere and shallow-atmosphere 
baroclinic instability, a balanced zonal flow, and a high-altitude orographic gravity wave verify the fidelity of the dynamics at low and high altitudes and for constant and variable composition atmospheres.  These results are compared to existing deep-atmosphere dynamical cores and a Fourier-ray code, indicating that the proposed discretized  equation sets are viable DyCore candidates for next-generation ground-to-thermosphere atmospheric models.
\end{abstract}

\end{frontmatter}
\newpage
\section{Introduction} \label{sec:intro}
Atmospheric dynamical cores (DyCores) form the foundation of modern numerical
weather prediction (NWP) and climate modeling. In providing accurate numerical
solutions to discretized forms of the atmospheric Euler equations, current
DyCores are developed primarily for the troposphere, the layer spanning
the lowest 10-15 km of the Earth’s atmosphere. Tropospheric skill improves
with increases in the spatial resolution of NWP models, yet gridpoint resolution is
constrained by finite operational computing resources. Thus considerable
ongoing effort is invested to develop robust, highly accurate,
fully compressible nonhydrostatic
atmospheric DyCores that exploit latest computational resources as efficiently
as possible to facilitate model runs at highest possible resolutions needed to improve
skill \citep[e.g.,][]{giraldo2010semi,ullrich2017dcmip2016,muelleretal2018}.

One routine approach to this computational challenge is to simplify or omit
terms in the governing equations that have negligible impacts on tropospheric
NWP. For example, since molecular viscosity and thermal conduction tendencies
have negligible impacts on tropospheric NWP at the space-time resolutions
currently possible using operational computing resources, these terms are
routinely omitted from DyCores. Small tropospheric molecular viscosities
allow turbulence to develop, and associated turbulent mixing that maintains a well-mixed troposphere
having a composition of $\sim$78\% $\mathrm{N_2}$, $\sim$21\% $\mathrm{O_2}$
and $\sim$1\% $\mathrm{Ar}$. This approximately
constant atmospheric mass composition in turn implies that the
atmosphere's mass specific heats at constant pressure, $c_p$, and at constant
volume, $c_v$, remain constants to a very good approximation from
$\sim$0--80~km altitude. Thus

\begin{equation}
 R = c_p - c_v , \label{eq:Rspec}
\end{equation}

\noindent
which interrelates atmospheric pressure $p$, density $\rho$ and temperature $T$ through
the ideal gas equation (a.k.a. equation of state),
\begin{equation}
 p = \rho R T , \label{eq:eos}
\end{equation}
can be set to a constant
($R =$~287.04~J~kg$^{-1}$~K$^{-1}$) to a very good approximation. 

Setting $R$ as a specific gas constant permits
considerable simplification of the governing equations, since, for example,
computing temperature and density from the prognostic energy and continuity
equations, respectively, immediately provides pressure diagnostically using
the equation of state with constant $R$, thus closing pressure gradient forces
in the momentum equations. Likewise, allowing the ratio of specific heats,

\begin{equation}
 \gamma = \frac{c_p}{c_v} , \label{eq:gamma}
\end{equation}

\noindent
to be a universal constant considerably simplifies the first law of
thermodynamics: it can be solved via a transport equation
for potential temperature $\theta=T/(p/p_0)^{\kappa}$, where $\kappa = R/c_p$.

Despite enormous increases in computing power and capacity, efficiency
concerns continue to motivate DyCore research and development as the scope and
computational complexity of forecasts expand. One such expansion motivating
the present work is a new need to extend upper boundaries to much higher
altitudes to support emerging space weather applications
\citep{akmaev2011,griffin2018numerical,griffin2018extension,jacksonetal2019}. The DyCore method can conceivably be
extended to so-called exobase altitudes at $\sim$500~km, above which
neutral particles may no longer behave as an ideal density-stratified fluid \citep[e.g.,][]{akmaev2011}.
The atmosphere above 100~km is known as
the thermosphere since daytime temperatures often exceed 1000~K (see Figure~\ref{fig:gasz}a)
due to absorption of intense solar radiation that dissociates $\mathrm{N_2}$ and
$\mathrm{O_2}$ molecules. This leads to substantial changes in thermospheric composition
with altitude, such that $R$ and $\gamma$ increase secularly with height from
$\sim$100--500~km rather than remaining constant
\citep[see Fig.~\ref{fig:gasz} as well as Fig.~1 of][]{eckermannetal2022}.
Thus standard NWP
DyCores incorporating inbuilt assumptions of constant $R$ and $\gamma$ become
highly inaccurate above 100~km altitude.

The challenge then is to design a modified or new DyCore that allows $R$
and $\gamma$ to vary arbitrarily in space and time, in order to capture
thermospheric dynamics accurately above 100~km, yet is still efficient and accurate
enough to yield comparable forecast skill and numerical efficiency
for operational NWP in the lower atmosphere below 100~km. Since
daytime thermospheric temperatures are up to 5 times larger than any encountered
in the troposphere and vary enormously between day and night (see Fig.~\ref{fig:gasz}a),
the
much larger wind and sound speeds that result present new challenges for stable
implicit-explicit (IMEX) time integration of the DyCore equations.
Atmospheric mass densities decrease by $\sim$13 orders of magnitude between
the surface (pressures $\sim$1000~hPa) and exobase pressures of
$\sim$10$^{-9}$~hPa near 500~km \citep{akmaev2011}, presenting substantial
new challenges for global mass conservation \cite{thuburn2008some} and numerical stability of
density- and pressure-dependent tendency terms in the discretized equations.

Given significant investments in existing NWP DyCores, one approach to these
challenges has been to modify existing NWP DyCores in ways that retain
pre-existing forms below 100~km but transition to modified thermospheric forms
above 100 km \citep[e.g.,][]{juang2011,eckermannetal2022}.
These modifications typically only work well when $R$ varies but $\gamma$
variations can be ignored, whereas including $\gamma$
variations exactly requires an additional prognostic equation for $\gamma$
\citep[or equivalently for $\kappa = R/c_p = 1-\gamma^{-1}$: e.g.,][]{juang2011,klemp2021,eckermannetal2022} that
adds new computational overhead and thus reduces efficiency for lower atmosphere NWP.
Thus here we pursue a more general and ambitious approach of reformulating to
a new DyCore that incorporates the variable $R$ and $\gamma$ needed for the thermosphere
yet retains requisite accuracy and efficiency to make it a viable
candidate for state-of-the-art tropospheric NWP as well.

In Section \ref{sec:eqset}, two versions of a variable composition DyCore based on specific internal energy are derived from the first law of thermodynamics and the continuity equation.  The pressure gradient and gravitational forces are formulated in a manner appropriate for thermospheric 
applications where density becomes highly rarefied.  In Section \ref{sec:numerics}, a mimetic spectral element method based on a hexahedral mesh is briefly described, along with the metric terms and time-integrators used in two atmospheric models: NUMA and NEPTUNE.  Section \ref{sec:laresults} shows numerical results from NUMA using the new DyCore developed below, including mass-conservation results.  High-altitude numerical results using the new DyCore in NEPTUNE
are shown in Section \ref{sec:haresults}.  Energy conservation is discussed in Section \ref{sec:energycons}, while conclusions are presented in Section \ref{sec:conclusion}.

\section{Ground-to-Exobase DyCore Equation Sets Incorporating Variable $R$ and $\gamma$}
\label{sec:eqset}
\subsection{NUMA and NEPTUNE}
Our development is formulated within the specific context of two atmospheric models
that solve deep-atmosphere forms of the fully compresssible nonhydrostatic
atmospheric Euler and Navier-Stokes equations using element-based Galerkin (EBG) methods
on tensor-product elements (quadrilaterals in 2D and hexahedra in 3D \cite{marras2016review,giraldo2020}).
The Nonhydrostatic Unified Model of the Atmosphere (NUMA) is a research-oriented
code that supports a range of experimental EGB DyCores and time integrators
\citep{giraldo2010semi,giraldoetal2013}.  NUMA serves as a DyCore testbed
environment for the Navy Environmental Prediction System Utilizing a Nonhydrostatic
Engine (NEPTUNE). NEPTUNE \citep{zaronetal2022}, currently under development to be the Navy’s next generation
operational NWP model, solves the fully compressible deep-atmosphere Euler equations using a spectral element method (SEM) on a global cube-sphere grid \cite{sadourny1972conservative}.  

\subsection{Current DyCores}
\cite{giraldo2010semi} presented a set of five separate forms of the fully compressible nonhydrostatic Euler equations for EGB discretization and numerical solution in NUMA as DyCores.
After a range of NUMA and early NEPTUNE experimentation, the inaugural NEPTUNE NWP DyCore has been
formulated using the $\theta$-based, vector-invariant equation set \cite{gardner2018implicit}, with small modifications from their equation set (2.3) to use
an Exner-function formulation for pressure gradient forces in the momentum equations.

However, this Set2NC set and all but one of the other NUMA equation sets presented in
\cite{giraldo2010semi} were derived on the inbuilt assumption that $R$ and $\gamma$
are both global constants.
The sole exception is the Set3C equation set (2.5) of
\cite{giraldo2010semi} that solves the equation
\begin{equation}
{{\partial E_{\mathrm{tot}}} \over {\partial t}} + {\grad} \cdot
\left[ \left(E_{\mathrm{tot}} + p\right) \mathbf{u} \right] =
	\mathfrak{D}(Q,\mathbf{D}) , \label{eq:tote}
\end{equation}
for the total energy density $E_{\mathrm{tot}}$, the sum of internal, kinetic
and potential energy density components
\begin{subequations}
\label{eq:etot0}
\begin{align}
	E_{\mathrm{tot}} &= E_i + E_k + E_p , \label{eq:etot1} \\
	                 &= \rho \left( e_i + e_k + \Phi \right) . \label{eq:etot2}
\end{align}
Here $t$ is time, $\mathbf{\grad}$ is the 3-component spatial gradient operator,
$\mathbf{u}$ is wind velocity, $\mathfrak{D}(Q,\mathbf{D})$ is a diabatic
term that depends on the diabatic heating rate $Q$ and the
mechanical drag/diffusion vector $\mathbf{D}$, 
\begin{equation}
 \Phi(z) = \int_{0}^{z} g(z') \, dz' , \label{eq:phi}
\end{equation}
is geopotential, $z$ is the height about sea level,
\begin{equation}
 e_i = c_v T , \label{eq:ei}
\end{equation}
is the specific internal energy (or internal energy per unit mass), and
\begin{equation}
 e_k = \frac{\mathbf{u} \cdot \mathbf{u}}{2} , \label{eq:ek}
\end{equation}
is specific kinetic energy.
\end{subequations}

Although Set3C is used in some atmospheric \cite{waruszewski2021entropy,souza2023}, climate \cite{sridhar2021large} 
and high altitude wave-propagation \cite{snively2008} models, 
it was not selected as the initial NEPTUNE DyCore. Thus we do not pursue a Set3C implementation here either.
Instead, we seek a new temperature-based equation set to implement and test in
NUMA that retains closer connections to the initial Set2NC implementation in NEPTUNE
but does not restrict to constant $R$ and $\gamma$.
We will show in section 7 that our final discretized equation set is
fully consistent with conservation of total energy density via
\eqref{eq:tote}--\eqref{eq:etot0}, while next sections show
that using the continuous equations.

\subsection{Internal Energy Equations}
The total energy equation \eqref{eq:tote}--\eqref{eq:etot0} can be reformulated without
approximation via the equations of state and mass continuity
into an equivalent form that solves for $E_i$, viz
\citep[e.g.,][]{akmaevjuang2008}
\begin{equation}
{{\partial E_{i}} \over {\partial t}} + \grad \cdot
	\left(E_{i} \mathbf{u} \right) + p  \grad \cdot \mathbf{u} =
	Q . \label{eq:ie1}
\end{equation}
This form has immediate advantages over the $E_{tot}$ forms in \eqref{eq:tote}--\eqref{eq:etot1}
in that, via \eqref{eq:etot1}--\eqref{eq:ei}, $E_i$ depends only on $\rho$, $c_v$
and $T$, and the nonconservative term on the right of \eqref{eq:ie1} involves only the
diabatic heating rate $Q$. However, since $\rho$ decreases by $\sim$13 orders of magnitude
from the surface to the exobase, $E_i$ will have a similarly huge range of values, presenting
potential challenges for discretization error and total energy conservation in
using $E_i$ as a prognostic variable.

Fortunately, it is straightforward to combine \eqref{eq:ie1} with the continuity
relation and equation of state to yield the equivalent form,
\begin{equation}
{{\partial e_i} \over {\partial t}} + \mathbf{u} \cdot
	{ \grad} e_i + \left( \gamma-1 \right) e_i
	{ \grad} \cdot \mathbf{u} =  \frac{Q}{\rho} , \label{eq:ie2}
\end{equation}
that uses specific internal energy $e_i = E_i/\rho = c_v T$ as the prognostic variable.

\begin{figure}[t] 
   \centering
   \includegraphics[width=0.99\textwidth]{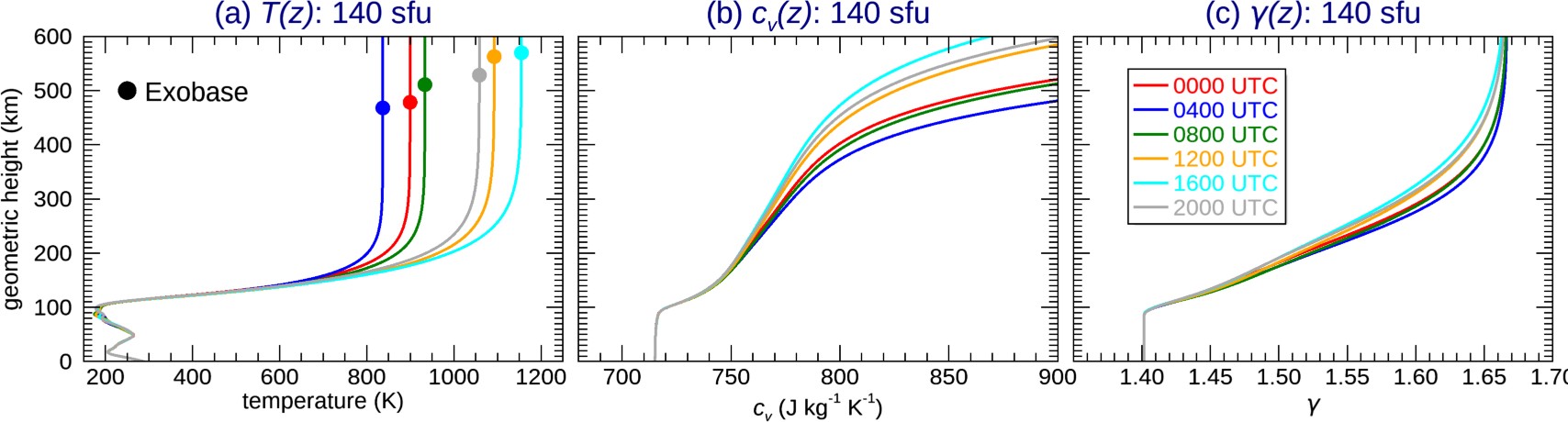}
   \caption{Global mean profiles of (a) $T(z)$, (b) $c_v(z)$, and (c) $\gamma(z)$
   evaluated using the MSIS model \citep{emmertetal2020}
   on day~80 using an intermediate value of the 10.7~cm solar radio flux
   of 140~sfu and a geomagnetic $A_p$ index of 4.
   Colored curves show results at six equispaced local times spanning a diurnal cycle.
   Filled circles in left column show exobase altitudes, computed
   using eq.~(3) of \cite{akmaev2011} with an O-O collisional cross section
   of 6~10$^{-15}$~cm$^2$ \cite[Fig.~2]{kharchenkoetal2000}.}
   \label{fig:gasz}
\end{figure}

\subsubsection{Advantages of $e_i$ for Ground-to-Exobase Prediction}
Using $e_i$ rather than $E_i$ as the prognostic energy variable removes
the density dependence and hence the large decrease in value between the
bottom and top of the domain. Via \eqref{eq:ei} and
Figure~\ref{fig:gasz} we see that $e_i$ varies
by no more than a factor of $\sim$6 between the surface
($c_v \sim$717~J~kg$^{-1}$~K$^{-1}$, $T \sim$250--300~K) and
$\sim$500~km altitude
($c_v \sim$900~J~kg$^{-1}$~K$^{-1}$, $T \sim$1200~K).

In this sense, replacing $E_i$ with $e_i$ is loosely analogous to using
$T$ instead of $\theta$
as a prognostic variable, since $\theta$ increases exponentially with
height. However, use of $e_i$ as a prognostic variable
does not require an additional prognostic tendency equation for
$\gamma$ (or equivalently $\kappa = 1 - \gamma^{-1}$) nor any virtualization
to account for variable $R$, both of which can be required
in general when using $T$ or $\theta$
\citep[e.g.,][]{klempwilhelmson1978,juang2011,klemp2021,eckermannetal2022}. The $c_v$ and
$\gamma$ values in \eqref{eq:ie2} are the exact local values derived
from standard diagnostic formulas based on the local composition environment
\citep[e.g., eqs. (1), (6) and (18)-(20) of][]{eckermannetal2022}. These
values can vary arbitrarily in space and time without any need to quantify their
local time tendencies or spatial gradients, while keeping \eqref{eq:ie2}
exact, thus providing exact $e_i$ solutions. Exact temperature solutions
follow by dividing this exact $e_i$ solution by the local $c_v$.

Indeed, \eqref{eq:ie2} is an exact form of the first law
of thermodynamics, as is illustrated by reformulating \eqref{eq:ie2} exactly and
equivalently into the familiar tendency form of the first law of thermodynamics,
\begin{subequations}
\label{eq:firstlaw}
\begin{equation}
 \frac{d e_i}{d t} + p \frac{d \alpha}{d t} = T \frac{ds}{dt} , \label{eq:ie3}
\end{equation}
where
\begin{equation}
 \frac{d}{dt} = {{\partial} \over {\partial t}} + \mathbf{u} \cdot {\grad}
 \label{eq:ddt}
\end{equation}
\end{subequations}
is the material derivative, $\alpha = \rho^{-1}$ is specific volume and $s$
is entropy, such that the heating rate $Q = (\rho/T) ds/dt$ involves nonconservation
of entropy due to external energy inputs (e.g., solar heating) or losses (e.g., radiative cooling).
Comparing \eqref{eq:ie2} and \eqref{eq:ie3} shows that the last term on the left-hand side of
\eqref{eq:ie2} is the familiar work term $p (d\alpha/dt)$.

The first law of thermodynamics is a fundamental relation that continues to hold generally
even as the thermosphere becomes progressively more rarefied
and disturbed by solar energy inputs with increasing height. For example,
at very high altitudes thermospheric density might be more usefully modelled
in certain circumstances in terms of
continuity relations for individual chemically-interacting major
species having individual production and loss terms, with the sum of these component
densities comprising the total thermospheric mass density \citep[e.g.,][]{dickinsonetal1984}.
Since $e_i$ has no explicit density dependence via \eqref{eq:ei},
it is unaffected directly by such details.
Furthermore, the specific internal energies of each individual major constituent can be
evaluated via
\eqref{eq:ei} using a specific $c_v$ for each constituent, such that the bulk
$e_i$ is given as the mixing-ratio-weighted sum of the specific internal energies of
each species, which in turn are collectively
governed by the first law of thermodynamics, and hence our $e_i$
equation \eqref{eq:ie2}: for details, see, for example, section 2 of
\cite{akmaevjuang2008} and references therein. Thus the $e_i$ equation \eqref{eq:ie2} holds
very generally in rarefied thermospheres of variable mass composition.
\subsection{Continuous Forms of Proposed DyCore Equations}
\subsubsection{Continuity Equation}
We consider two equivalent forms of the continuity equation,
\begin{subequations}
\label{eq:conts}
\begin{equation}
  \frac{\partial \rho}{\partial t} + \mathbf{u} \cdot {\grad} \rho + \rho { \grad} \cdot \mathbf{u} = 0 ,\label{eq:cont1}
\end{equation}
\begin{equation}
  \frac{\partial \rho}{\partial t} + { \grad} \cdot \left( \rho \mathbf{u} \right) = 0 , \label{eq:cont2}
\end{equation}
\end{subequations}
related via the product rule (PR) of differentiation.
Since the PR is not necessarily satisfied
exactly when using discretized divergence and gradient operators, we treat the two forms
in \eqref{eq:conts} as separate entities. We refer to \eqref{eq:cont1} as
the PR form since it provides $\partial \rho/\partial t$ as the two term PR expansion of
${\grad} \cdot ( \rho \mathbf{u} )$, and thus to \eqref{eq:cont2}
as the no-PR (or flux) form.

\subsubsection{Specific Internal Energy Equation}
Since the prognostic $e_i$ equation \eqref{eq:ie2} follows from the first law
of thermodynamics \eqref{eq:ie3} using the PR form \eqref{eq:cont1} of the continuity equation,
we refer to it as the PR form
of the specific internal energy equation. The
corresponding no-PR form, derived from \eqref{eq:ie3} using \eqref{eq:cont2}, is
\begin{equation}
{{\partial e_i} \over {\partial t}} + \mathbf{u} \cdot
	{\grad} e_i + \frac{\left( \gamma-1 \right) e_i}{\rho} \left[
	{\grad} \cdot \left( \rho \mathbf{u} \right) - \mathbf{u} \cdot { \grad} \rho \right]
	=  \frac{Q}{\rho} . \label{eq:ienopr}
\end{equation}

\subsubsection{Equation of State}
Using $e_i$ as the prognostic variable, the equation of state \eqref{eq:eos} takes
the form
\begin{equation}
 p = \rho \left( \gamma - 1 \right) e_i . \label{eq:eosei}
\end{equation}

\subsubsection{Momentum Equations}
The momentum equations take the usual general form
\begin{equation}
{{\partial \mathbf{u}} \over {\partial t}} + \mathbf{u} .
	{\grad} \mathbf{u} + 2 \mathbf{\Omega} \times \mathbf{u} - \mathbf{f}_{\mathrm{pg}}
	+ \grad \Phi  = \rho^{-1} \mathbf{D} , \label{eq:momeq}
\end{equation}
where $\mathbf{\Omega}$ is the Earth's angular velocity, $\rho \mathbf{f}_{\mathrm{pg}}$
is the pressure gradient force, $\grad \Phi = \mathbf{g}$ is the geopotential,
and $\mathbf{D}$ is the drag/diffusion force.

\paragraph{Pressure Gradient Force}
The pressure gradient force per unit mass is
\begin{equation}
 \mathbf{f}_{\mathrm{pg}} = - \frac{{ \grad}p}{\rho} , \label{eq:pgf1}
\end{equation}
where pressure is computed diagnostically from the equation of state \eqref{eq:eosei}.

Since $\rho$ and $e_i$ in \eqref{eq:eosei} are prognostic variables represented by polynomial
basis functions, aliasing will occur once the gradient is discretized using the SEM.
Since $\gamma$ also varies through the thermosphere (Fig.~\ref{fig:gasz}c), the diagnostic
pressure derived from \eqref{eq:eosei} possesses a cubic nonlinearity that accentuates aliasing
error. Although the advective terms in \eqref{eq:momeq} also possess aliasing nonlinearities,
the $\rho^{-1}$ dependence of \eqref{eq:pgf1} will cause any errors to grow with height and
become large at high altitudes.

We circumvent this problem by inserting \eqref{eq:eosei} into \eqref{eq:pgf1}
and expanding via the PR, such that
\begin{subequations}
\label{eq:pgf2}
\begin{align}
 - \mathbf{f}_{\mathrm{pg}} = \frac{{\grad}p}{\rho}
&= \left(\gamma-1\right) e_i \frac{{ \grad}\rho}{\rho} +
   \left(\gamma-1\right) {\grad}e_i + e_i { \grad} \gamma , \label{eq:pgf2a} \\
&= \left(\gamma-1\right) e_i {\grad}\log \rho +
   \left(\gamma-1\right) { \grad}e_i + e_i { \grad} \gamma . \label{eq:pgf2b}
\end{align}
\end{subequations}
The modified expression \eqref{eq:pgf2b} has several advantages over \eqref{eq:pgf1}.
First, the amplifying $\rho^{-1}$ term in \eqref{eq:pgf1} is removed. Second, there
is no need to evaluate a diagnostic $p$ via \eqref{eq:eosei}, with \eqref{eq:pgf2b}
depending on the prognostic variables $\rho$ and $e_i$ and the local $\gamma$.
These simplifications come at the expense of having to compute a new ${ \grad} \gamma$
term explicitly.

\paragraph{Gravitational Force}
Given extension to exobase altitudes using deep-atmosphere formulations of the equations,
we include height-dependent gravitational acceleration
in the gravitational force $-\rho \mathbf{g}$ but ignore small horizontal variations
\citep[as considered, e.g., by][]{whiteetal2008}, such that
\begin{subequations}
\label{eq:gravacc}
 \begin{equation}
  \mathbf{g} = g(z) \mathbf{\hat{r}} , \label{eq:gvec}
 \end{equation}
 \begin{equation}
  g(z) = \frac{g_0}{\left( 1 + z/a \right)^2} , \label{eq:gz}
 \end{equation}
\end{subequations}
where $z$ is height, $a$ is Earth radius, $\mathbf{\hat{r}}$ is the unit radial vector, and $g_0 =$~9.8066~m~s$^{-2}$ is surface
gravitational acceleration. As discussed in section~3.2, we find superior numerical
performance when the gravitational acceleration is formulated as the discrete gradient
of the geopotential. Combining
\eqref{eq:phi} and \eqref{eq:gz} yields the analytical solution
\begin{equation}
  \Phi(z) = \frac{g_0 z}{1 + z/a} . \label{eq:Phian}
\end{equation}
Note that in the shallow-atmosphere limit $z/a \rightarrow 0$, so that
$\partial \Phi/\partial z \rightarrow g_0$
in \eqref{eq:Phian}.

\paragraph{Coriolis Force}
The Coriolis force $-2 \rho \mathbf{\Omega} \times \mathbf{u}$ is included without any
approximations, where $\mathbf{\Omega} = \hat{\Omega} \mathbf{\hat{k}}$ and
$ \mathbf{\hat{k}}$ is a unit vector aligned
with the axis of the earth.

\subsection{Complete Equation Sets}
We summarize the continuous PR and no-PR forms of the equations sets we propose as
the basis of a ground-to-exobase DyCore.
\subsubsection{PR Form}
The PR set combines \eqref{eq:ie2}, \eqref{eq:cont1}, \eqref{eq:eosei} and \eqref{eq:momeq}:
\begin{subequations}
\label{eq:set4ncpr}
\begin{equation}
 \frac{\partial \rho}{\partial t} + \mathbf{u} \cdot { \grad} \rho + \rho { \grad} \cdot \mathbf{u} = 0 , \label{eq:comasspr}
\end{equation}
\begin{equation}
{{\partial \mathbf{u}} \over {\partial t}} + \mathbf{u} \cdot
	{\grad} \mathbf{u} + 2 \mathbf{\Omega} \times \mathbf{u} - \mathbf{f}_{\mathrm{pg}}
	+ \grad \Phi = \rho^{-1} \mathbf{D} , \label{eq:mompr}
\end{equation}
\begin{equation}
{{\partial e_i} \over {\partial t}} + \mathbf{u} \cdot
	{\grad} e_i + \left( \gamma-1 \right) e_i
	{ \grad} \cdot \mathbf{u} =  \rho^{-1} Q , \label{eq:eipr}
\end{equation}
\begin{equation}
 p = \rho \left( \gamma - 1 \right) e_i . \label{eq:eospr}
\end{equation}
\end{subequations}
\subsubsection{No-PR Form}
The no-PR set combines \eqref{eq:cont2}, \eqref{eq:ienopr}, \eqref{eq:eosei}
and \eqref{eq:momeq}:
\begin{subequations}
\label{eq:set4ncnopr}
\begin{equation}
 \frac{\partial \rho}{\partial t} + { \grad} \cdot \left( \rho \mathbf{u} \right) = 0 , \label{eq:comassnopr}
\end{equation}
\begin{equation}
{{\partial \mathbf{u}} \over {\partial t}} + \mathbf{u} \cdot
	{ \grad} \mathbf{u} + 2 \mathbf{\Omega} \times \mathbf{u} - \mathbf{f}_{\mathrm{pg}}
	+ \grad \Phi = \rho^{-1} \mathbf{D} , \label{eq:momnopr}
\end{equation}
\begin{equation}
{{\partial e_i} \over {\partial t}} + \mathbf{u} \cdot
	{\grad} e_i + \frac{\left( \gamma-1 \right) e_i}{\rho} \left[
	{\grad} \cdot \left( \rho \mathbf{u} \right) - \mathbf{u} \cdot { \grad} \rho \right]
	=  \rho^{-1} Q . \label{eq:einopr}
\end{equation}
\begin{equation}
 p = \rho \left( \gamma - 1 \right) e_i . \label{eq:eosnopr}
\end{equation}
\end{subequations}

\section{Numerical Methods}
\label{sec:numerics}
\subsection{Mimetic Spectral Element Method (SEM)}
\label{sec:sem}
The proposed continuous equation sets \eqref{eq:set4ncpr}-\eqref{eq:set4ncnopr} were implemented in discretized form within our two spectral element (SE) models.  In this section, we give a brief outline of the SEM spatial discretization used for the PR form \eqref{eq:set4ncpr} and no-PR form \eqref{eq:set4ncnopr} of the DyCore.  For additional details, the reader may consult \cite{giraldo2020} or \cite{kelly2012continuous}.  

Both NUMA and NEPTUNE use a mimetic SEM that decomposes the domain of interest $\Omega$ into a collection of $N_e$ non-overlapping elements $\Omega^e$.
A \emph{compatible} (or \emph{mimetic}) SEM mimics fundamental vector calculus identities in a discrete sense \cite{taylorfournier2010}.  For example, consider \emph{Green's identity} for sufficiently smooth $\varphi$ and $\mathbf{u}$
\begin{equation}
\int_{\Omega_e} \mathbf{u} \cdot \grad \varphi \, d\Omega_e + \int_{\Omega_e} \varphi \grad \cdot \mathbf{u} \, d\Omega_e = \int_{\Gamma_e} \varphi \mathbf{u} \cdot \mathbf{\hat{n}} \, d\Gamma_e ,
\label{eq:green1}
\end{equation}
where $\mathbf{\hat{n}}$ is the outwardly facing unit normal vector and $\Gamma_e$ is the boundary of $\Omega_e$.  A discrete analog of \eqref{eq:green1}
is given by
\begin{equation*}
\langle \mathbf{u} \cdot \grad_d f \rangle_{\Omega_e} + \langle f \grad_d \cdot \mathbf{u} \rangle_{\Omega_e} = \langle f \mathbf{u} \cdot \mathbf{\hat{n}} \rangle_{\Gamma_e} ,
\label{discdiv}
\end{equation*}
where $\langle \cdot \rangle$ denotes discrete (inexact) integration and ${ \grad}_d$ is the discrete form of ${ \grad}$. Inexact integration, which uses quadrature points that are coincident with interpolation nodes,
is chosen in order to (a) produce a diagonal mass matrix and (b) reduce the computational complexity of the inviscid operators to $\mathcal{O} \left(N_e n^4 \right)$, where $n$ is the polynomial order.  
The discrete divergence and gradient operators on an element $\Omega^e$ are defined via
\begin{subequations}
\label{discops}
\begin{equation}
\grad_d \cdot \mathbf{u} = \frac{1}{J} \sum_{\alpha=1}^3 \frac{\partial \left( J u^{\alpha} \right)}{\partial x^{\alpha}}
\label{discdiv}
\end{equation}
\begin{equation}
\grad_d f = \sum_{\alpha=1}^3 \frac{\partial f}{\partial x^{\alpha}} \mathbf{\hat{e}}^{\alpha}
\label{discgrad}
\end{equation}
\end{subequations}
where $u^{\alpha}$ are the three contravariant components of velocity, $\mathbf{\hat{e}}^{\alpha}$ are the contravariant basis vectors that are oriented normally with respect to each element, and $J > 0$ is the determinant of the metric Jacobian matrix.  For further details, the reader may 
consult section 6.2 in \cite{kopriva2009implementing}.

Let $\psi_i (\mathbf{x})$ be a basis/test function constructed as a tensor product of Lagrange polynomials. Following \cite{abdi2016efficient}, define the 
finite dimensional function space
\begin{equation*}
\mathcal{V}^{CG}_N = \left\{ \psi \in H^1 \left( \Omega \right) : \psi \in \mathbb{P}_N \left( \Omega_e \right) \right\} ,
\label{cgspace}
\end{equation*}
where $ H^1 \left( \Omega \right) \subset C^0 \left( \Omega \right)$ and $\mathbb{P}_N$ is the space of $N$-th order polynomials.  The continuity between elements $\Omega_e$ is enforced via the direct stiffness
summation (DSS) operator.  We consider two categories of weak-forms illustrated via the PR and no-PR forms of the continuity equation \eqref{eq:comasspr} and the no-PR continuity equation \eqref{eq:comassnopr}, respectively.  In the first case, termed the \emph{strong variational weak form}, we merely multiply $\psi_i (\mathbf{x})$ by the governing equation and integrate over the domain $\Omega$ with boundary $\Gamma$.  This is the weak form utilized in the NWP version of  NEPTUNE \cite{zaronetal2022}.  In the second case, termed the \emph{weak weak form}, an integration by parts is performed, yielding
\begin{equation}
\int_{\Omega} \left[ \psi_i \frac{\partial \rho}{\partial t} - \grad_d \psi_i \cdot \left( \rho \mathbf{u} \right) \right] \, d\Omega + \int_{\Gamma} \psi_i \rho \mathbf{u} \cdot \mathbf{\hat{n}} \, d\Gamma = 0 .
\label{eq:weakweak}
\end{equation}
In all our tests, we enforce a no-mass flux (or rigid) boundary condition (BC) on both the lower (ground) and upper boundaries where the normal velocity $\mathbf{u} \cdot \mathbf{\hat{n}}$ is set to zero.  Hence, the surface integral in \eqref{eq:weakweak} vanishes as a result of imposing the rigid BC in a \emph{weak sense}.  In contrast, the rigid BCs must be enforced in a \emph{strong sense} for the strong weak form by altering the prognostic vertical velocity $w$. 
\subsection{Grid and Metric Terms}
\label{sec:gridmetrics}
Both NUMA and NEPTUNE use an equi-angular cubed sphere \cite{sadourny1972conservative, ronchi1996cubed} SE grid for each horizontal level expressed in Cartesian coordinates.  Let $\bm{\xi} = (\xi, \eta, \zeta) = \left(\xi_1, \xi_2, \xi_3 \right)$ be the local coordinates
within each reference element (cube), and let $\bm{x} = (x,y,z) = \left(x_1, x_2, x_3 \right)$ be the physical Cartesian coordinates of each physical element.  Grids are constructed using concentric cubed-sphere spherical shells with spherical coordinates $(r, \phi, \lambda)$, where $r$ is the distance from the center of the earth, $\phi$ is latitude, and $\lambda$ is longitude.  By this construction $r = r(\zeta)$ and $\zeta = \zeta(r)$, while $(\phi, \lambda) = F(\xi, \eta)$
and $(\xi, \eta) = F^{-1} (\phi, \lambda)$.  In NUMA, a Gal-Chen-Somerville terrain-following coordinate \cite{gal1975use} is employed in the vertical 
by warping the computational grid.  NEPTUNE incorporates a range of terrain-following vertical coordinates, and for all NWP-type runs uses a hybrid coordinate that transitions from terrain to pure height surfaces at some specifiable altitude that is not constrained to be the upper boundary.  


Each hexahedral element is mapped onto a reference computational element (cube) via the mapping $\bm{x} = \bm{X} \left( \bm{\xi} \right)$.  In addition, a source term representing height-dependent gravitational acceleration \eqref{eq:gravacc} must be constructed.  These metrics must be constructed carefully in order to prevent spurious source terms from degrading the solution, which can result in model instability.  Rigorous derivations in \cite{kopriva2006metric} and \cite{kopriva2019free} provide these metric terms and spurious noise terms from a DG viewpoint, while \cite[Chapter 6]{kopriva2009implementing} provides a corresponding CG analysis.  We conclude here by reviewing 3 different forms for the metric terms, each of which we test using NUMA in section 4.2.

\subsubsection{Cross-Product Form}

The standard method of computing the metric terms (the contravariant vectors) uses the cross-product of the covariant vectors resulting in Eq.\ (12.30) in \cite{giraldo2020} which are written as
\begin{equation}
\grad \xi^i = \frac{1}{J} \left( \diff{\vc{x}}{\xi^j} \times \diff{\vc{x}}{\xi^k} \right) ,
\label{eq:cross-product-metrics}
\end{equation}
where the determinant of the metric Jacobian is defined as $J=\diff{\vc{x} }{\xi} \cdot \left( \diff{\vc{x} }{\eta} \times \diff{\vc{x} }{\zeta} \right) $ and $i,j,k$ are defined cyclically such that if $i=1$, then $j=2$, and $k=3$, etc.  We refer to the metric terms given in \eqref{eq:cross-product-metrics} as the \emph{cross-product} form.  
Since both NUMA and NEPTUNE use Cartesian coordinates, the cross-product form may not represent the spherical domain accurately when coarse grids are used.  

\subsubsection{Semi-Analytic Form}

The \emph{semi-analytic} metrics \cite{nair2005} address a cross-product deficiency by building the spherical geometry into the metric terms.  We again use \eqref{eq:cross-product-metrics} but include the map from spherical to Cartesian coordinates, as follows:
\begin{equation}
x = r \cos \phi \cos \lambda; \, y = r \cos \phi \sin \lambda; \, z = r \sin \phi ,
\label{eq:spherical_to_cartesian}
\end{equation}
where $\phi \in [ -\frac{\pi}{2}, +\frac{\pi}{2}]$ and $\lambda \in [ 0, 2 \pi ]$ denote the latitude and longitude and $r$ is the radius. We can now use the chain rule as follows:
\begin{equation*}
\diff{\vc{x}}{\xi} = \diff{\vc{x}}{r} \diff{r}{\xi} + \diff{\vc{x}}{\phi} \diff{\phi}{\xi} + \diff{\vc{x}}{\lambda} \diff{\lambda}{\xi} ,
\end{equation*}
where terms such as $\diff{\vc{x}}{r}$ are computed using \eqref{eq:spherical_to_cartesian}.  This approach is ideal on spherical domains; however, it may not be accurate if steep topography is present.

\subsubsection{Curl Invariant Metrics}

\emph{Curl-invariant} (CI) metrics \cite{kopriva2006metric} are constructed to satisfy constant-state preservation, whereby a constant flow field remains unchanged as time evolves. 
This property is connected to well-balancing which aids in
satisfying a discrete hydrostatic balance \cite{berberich2021}.

These metrics are expressed as
\begin{equation}
\grad \xi^i = \frac{1}{2J} \left[   \diff{ }{\xi^k} \left( \diff{\vc{x}}{\xi^j} \times \vc{x} \right) - \diff{ }{\xi^j} \left( \diff{\vc{x}}{\xi^k} \times \vc{x} \right)  \right] ,
\label{eq:curl-invariant-metrics}
\end{equation}
where, once again, the indices $(i,j,k)$ are cyclic.  Unlike the semi-analytic metrics, the CI metrics do not make any \emph{a priori} assumptions about the grid or 
the presence of terrain. 
\subsection{Time-Integration (TI) Methods}
\label{sec:tis}
%
NUMA accomodates use of a range of IMplicit-EXplicit (IMEX) methods \cite{giraldoetal2013}, including Horizontally Explicit Vertically Implicit (HEVI) \cite{gardner2018implicit,alves2023} time-integration (TI) methods, while NEPTUNE relies solely on HEVI TI methods.  In linear IMEX methods, the individual adiabatic terms in the governing equations are typically first linearized.  The linear component is then solved 
implicitly in time, while the remaining terms are solved explicitly.  Physically, the fast waves are approximated using implicit linear solutions, while slower waves are solved explicitly without such approximations, thereby allowing a larger time step than a purely explicit method.  Alternatively, we can avoid linearization of the implicit terms, but this approach requires using a nonlinear solver (e.g., Newton's method) \cite{knoll2004jacobian}. The maximum time-step that maintains stability is constrained by the slow waves only.  NUMA incorporates flexible IMEX machinery \cite{giraldoetal2013} that can be used to construct efficient and flexible TI schemes.  Linear IMEX schemes may either use a static reference state or a dynamically updated reference state.  Both approaches are described below.

In the numerical experiments shown in section \ref{sec:laresults}, we use the second-order additive Runge-Kutta (ARK2b) TI described in \cite{giraldoetal2013}.  We have constructed two flavors of IMEX methods: 1) linearization over a fixed, hydrostatic reference state and 2) linearization over a previous time step.  The latter is referred to as linear-HEVI (L-HEVI) \cite{alves2023} since this scheme is equivalent to a single Newton iteration of a fully implicit
scheme or HEVI scheme \cite{gardner2018implicit, vogletal2019}.  The L-HEVI scheme is similar to Rosenbrock schemes \cite{ullrich2012operator} and does not require an
\emph{a priori} reference field, making it appropriate for ground-to-exobase applications where the thermosphere undergoes very large changes in temperature and density between the hot dayside and cold nightside.  Newton iterations are used to solve the nonlinear system of equations, and then a direct solve is used for the linear terms.  In particular, the Jacobian matrix is computed analytically, and LAPACK \cite{lapack99} is used to solve the system.  A Jacobian-free Newton-Krylov (JFNK) algorithm was also tested, but a direct solve was found to be much faster.  An extensive comparison is carried out in \cite{alves2023}.  
%
\section{Low-Altitude Numerical Experiments}
\label{sec:laresults}
Our DyCore must be efficient and accurate for both tropospheric NWP applications using a low-top configuration and ground-to-exobase NWP applications using a high-top configuration. We test the former here, and the latter in section \ref{sec:haresults}.
%
%
\subsection{Nonhydrostatic Baroclinic Instability without Terrain}
%
A popular test of deep-atmosphere, nonhydrostatic dynamical cores is the idealized Ullrich baroclinic instability (BI) experiment \cite{ullrich2014proposed}.  This case is included in the DCMIP 2016 suite \cite{ullrich2017dcmip2016} and has been used to validate numerous DyCores \cite{gardner2018implicit,skamarock2021fully,borchert2019upper,waruszewski2021entropy}.  We ran this test case in NUMA using our new SE-discretized DyCore equations with 24 elements per cube sphere panel and polynomial order $n=4$, which has an average horizontal resolution of 104 km at the equator and an equivalent angular resolution $\sim 0.94^{\circ}$ along the equator (104 km at the surface).  We used 33 vertical levels (8 SEs of $n=4$) with the grid stretching specified in \cite{ullrich2014proposed} with a model top of 30 km and a rigid boundary condition at both the lower and upper boundaries.  A second-order, one-dimensional IMEX TI method based on an additive Runge-Kutta (ARK2) scheme was used, in which the linearization was performed over the previous time-step \cite{alves2023}. 

NUMA was run using both the no-PR and PR forms of the equation sets \eqref{eq:set4ncpr} and \eqref{eq:set4ncnopr}.  A time-step of $\Delta t = $ 50.8235 s was used with a no-Schur formulation with the PR form.  This time-step may be increased to $\Delta t =$ 120 s with the no-PR form \eqref{eq:set4ncnopr} without introducing any instability.  Fourth-order horizontal hyperdiffusion with a constant coefficient of $9 \times 10^{14} \mbox{m}^4 \mbox{s}^{-1}$ was used to stabilize the dynamics.  No hyperdiffusion or explicit filtering was utilized in the vertical; rather, NUMA relies on the implicit diffusion provided by the 1D IMEX TI method to dissipate grid-point noise in the vertical.  

\begin{figure}
\centering
\subfigure[]{\includegraphics[width=3in]{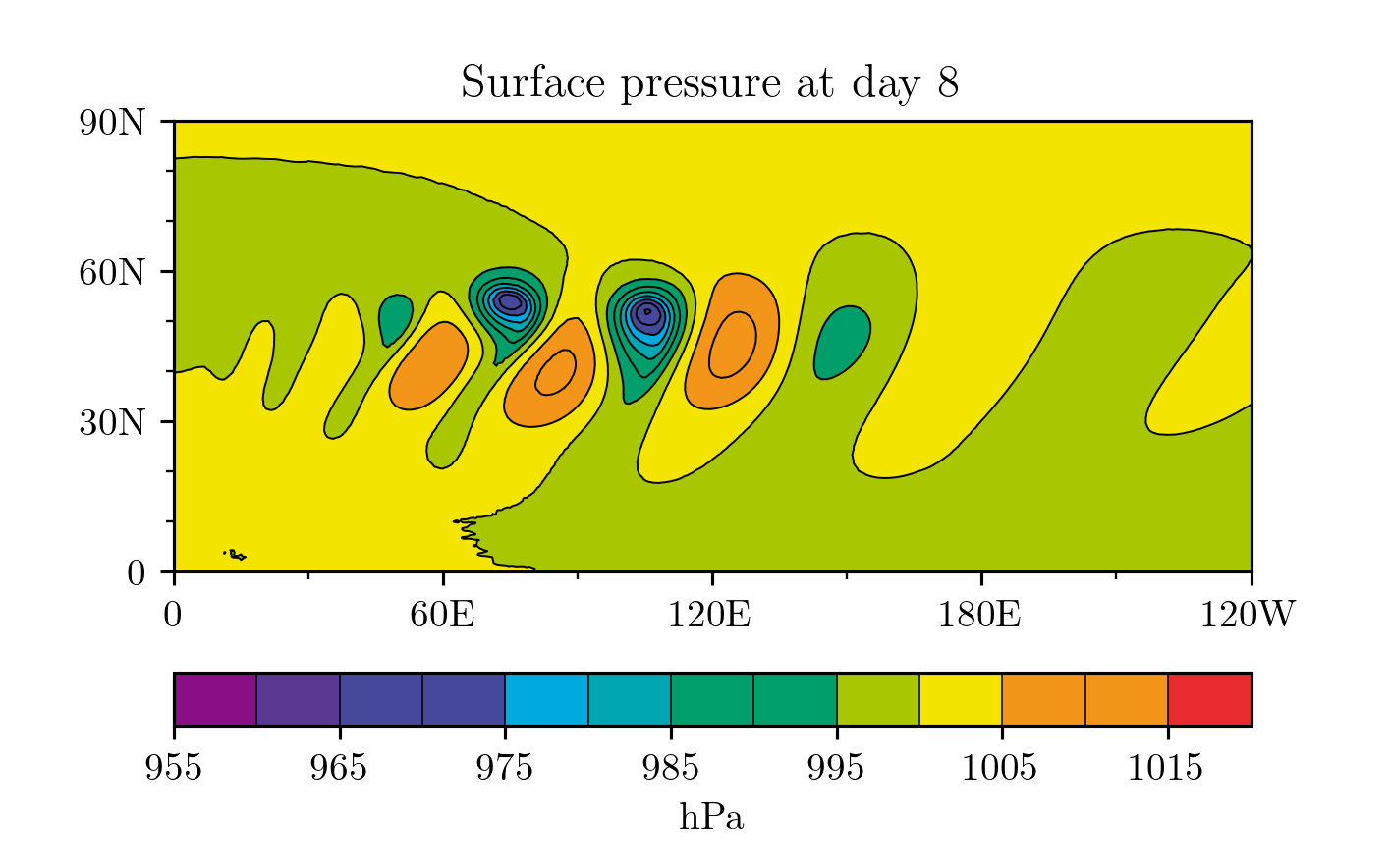}}
\subfigure[]{\includegraphics[width=3in]{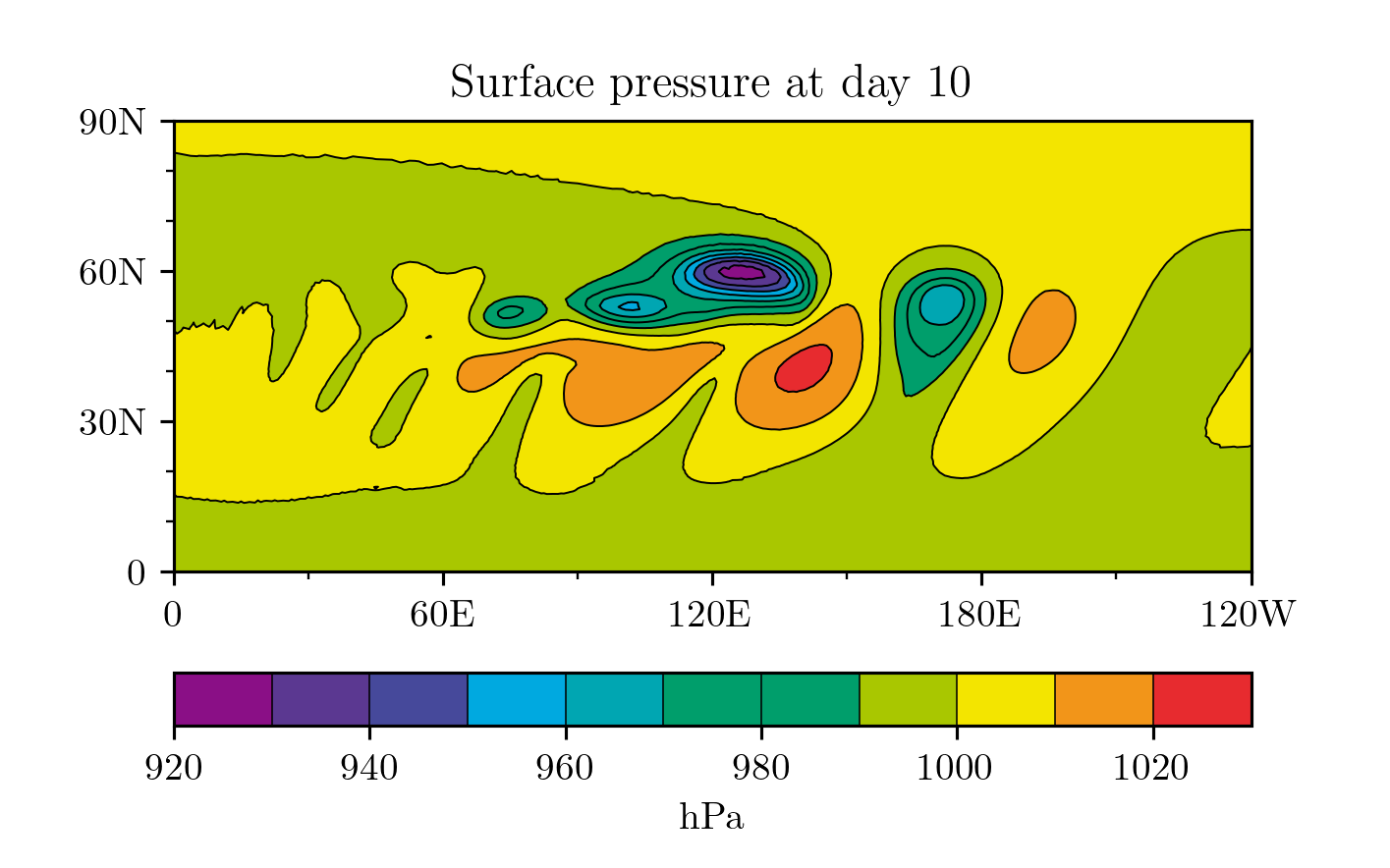}}
\subfigure[]{\includegraphics[width=3in]{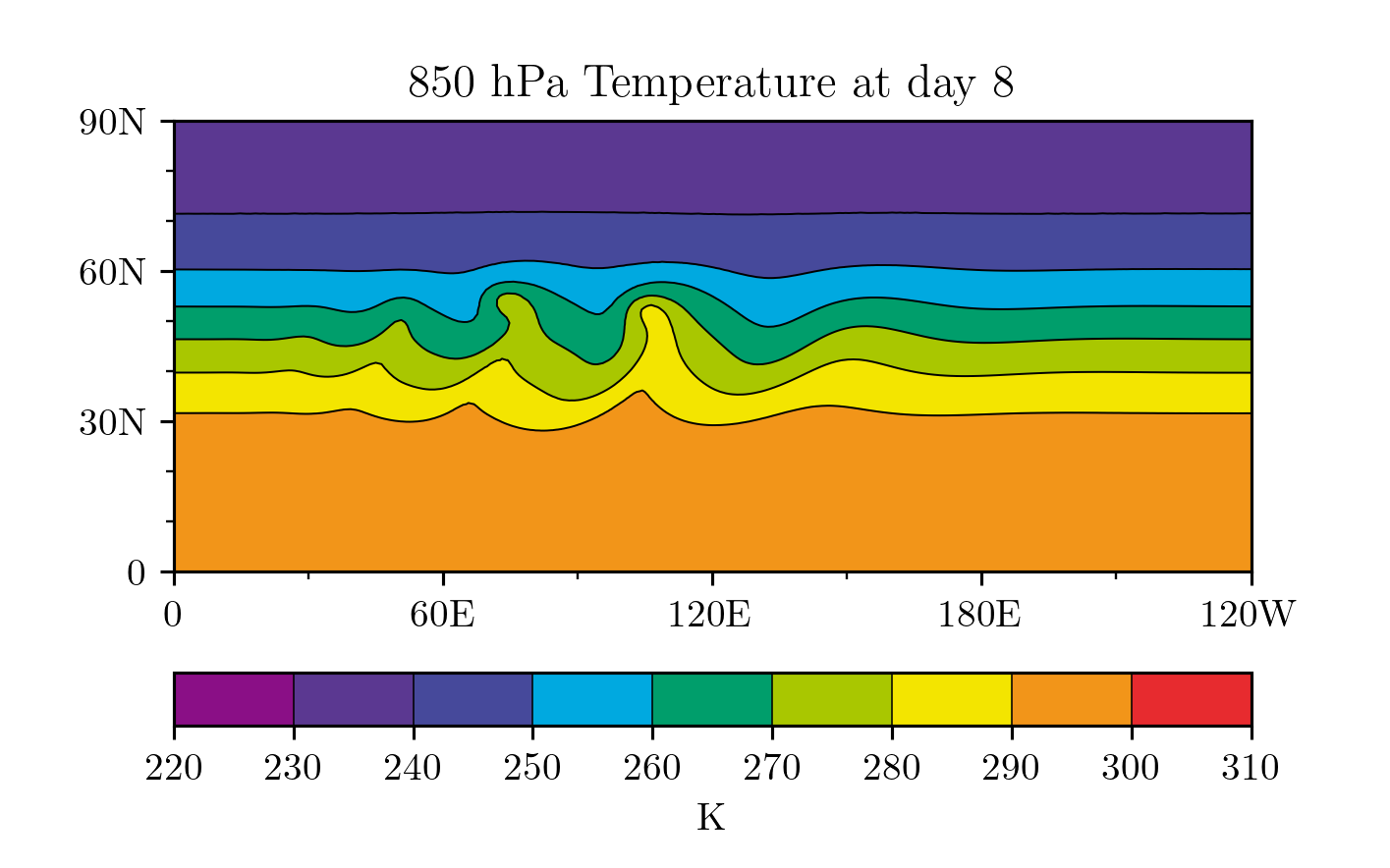}}
\subfigure[]{\includegraphics[width=3in]{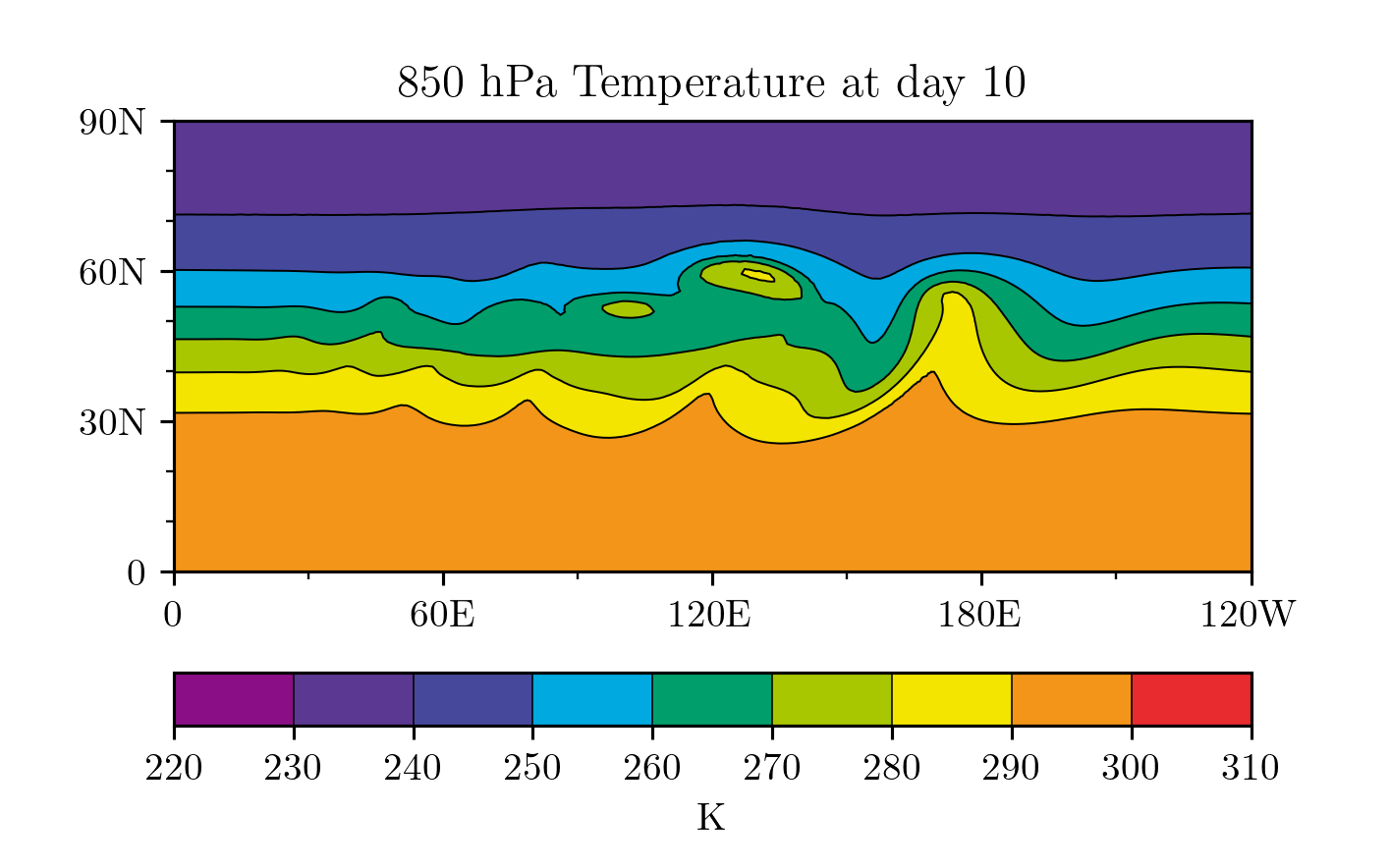}}
\subfigure[]{\includegraphics[width=3in]{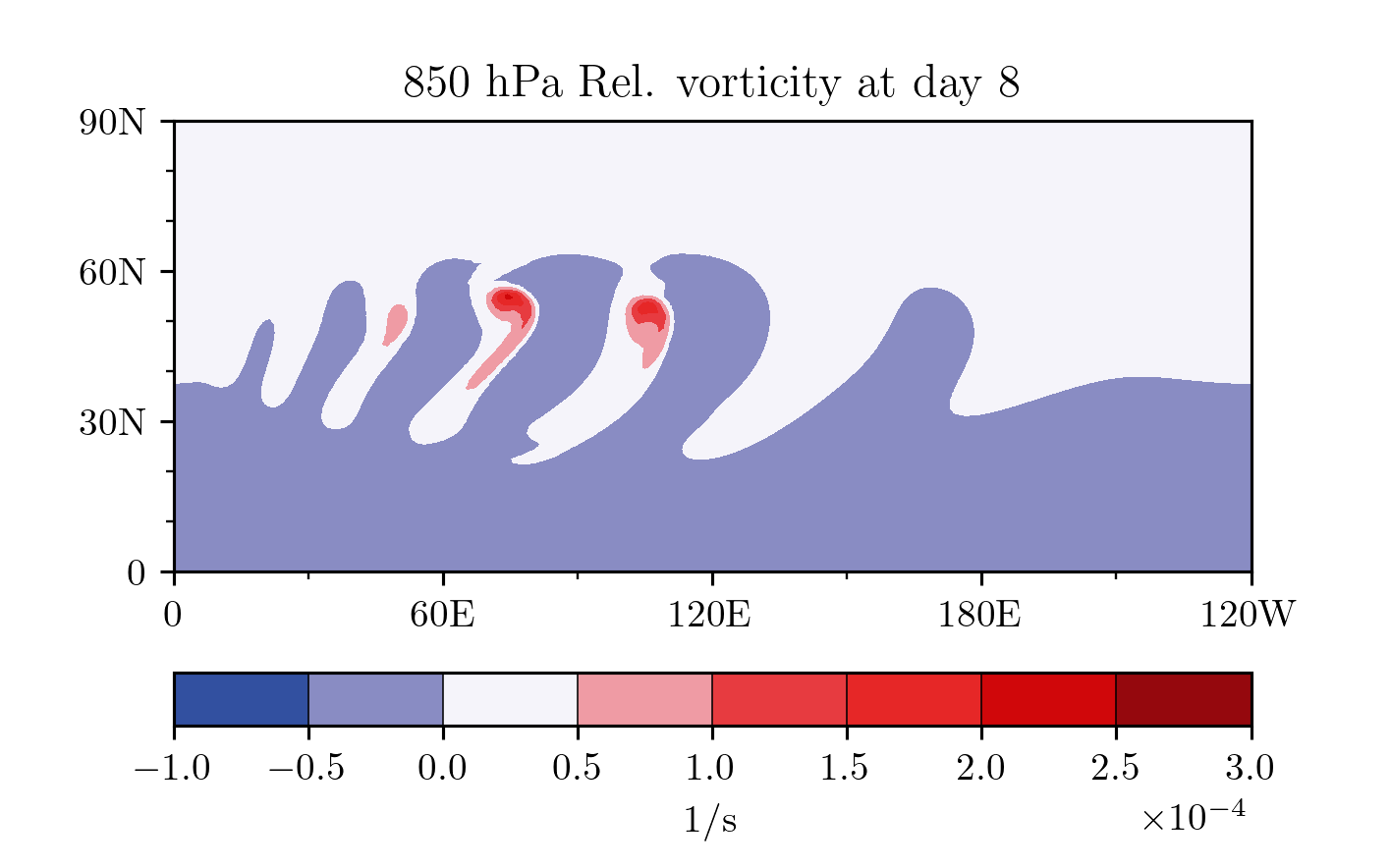}}
\subfigure[]{\includegraphics[width=3in]{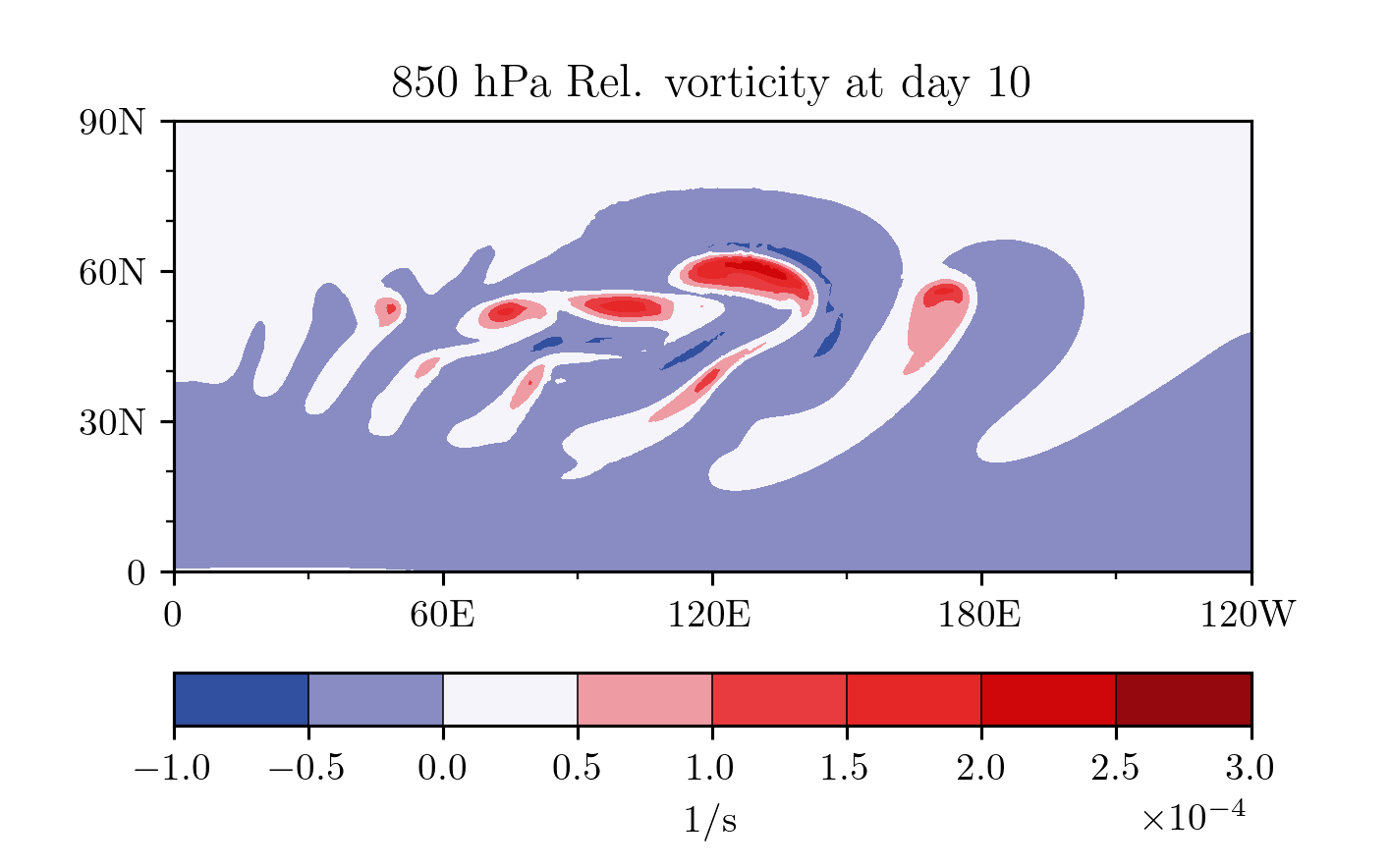}}
\caption{Surface pressure, 850 hPa temperature, and 850 hPa vorticity for day 8 (left) and day 10 (right) for NUMA using the no-PR formulation.}
\label{fig:case600days8and10}
\end{figure}

Figure \ref{fig:case600days8and10} displays the surface pressure, 850 hPa temperature, and 850 hPa vorticity as a function of latitude and longitude using the no-PR form after 8 and 10 days of NUMA integration.  Corresponding results for the PR form are virtually indistinguishable except for the surface pressure at day 10.  In the no-PR form, the contours of surface pressure are noisier, indicating that there is less implicit diffusion near the surface in the no-PR form due to the weak imposition of the rigid BC.  In addition, the surface pressure and 850 hPa contour plots in Fig.~\ref{fig:case600days8and10} agree well with the corresponding results for the MCore (Fig.~4 in \cite{ullrich2014proposed}), ENDGame (Fig.~5 in \cite{ullrich2014proposed}), and MPAS (Fig.~2 in \cite{skamarock2021fully}) DyCores.  There are small observable differences in 850 hPa vorticity, indicating that this diagnostic is very sensitive to the choice of discretization, stabilization, and filtering.  

As additional verification, we ran this same experiment using a separate code called Atum incorporating an entropy-stable DG method \cite{waruszewski2021entropy,souza2023} using the same spatial resolution as NUMA.  Atum uses the conservative form of the total energy equation given by \eqref{eq:tote}--\eqref{eq:etot0}, which conserves both mass and total energy when used with a consistent DG method and a no-mass flux BC.  

\begin{figure}
\centering
\subfigure[minimum surface pressure]{\includegraphics[width=3.0in]{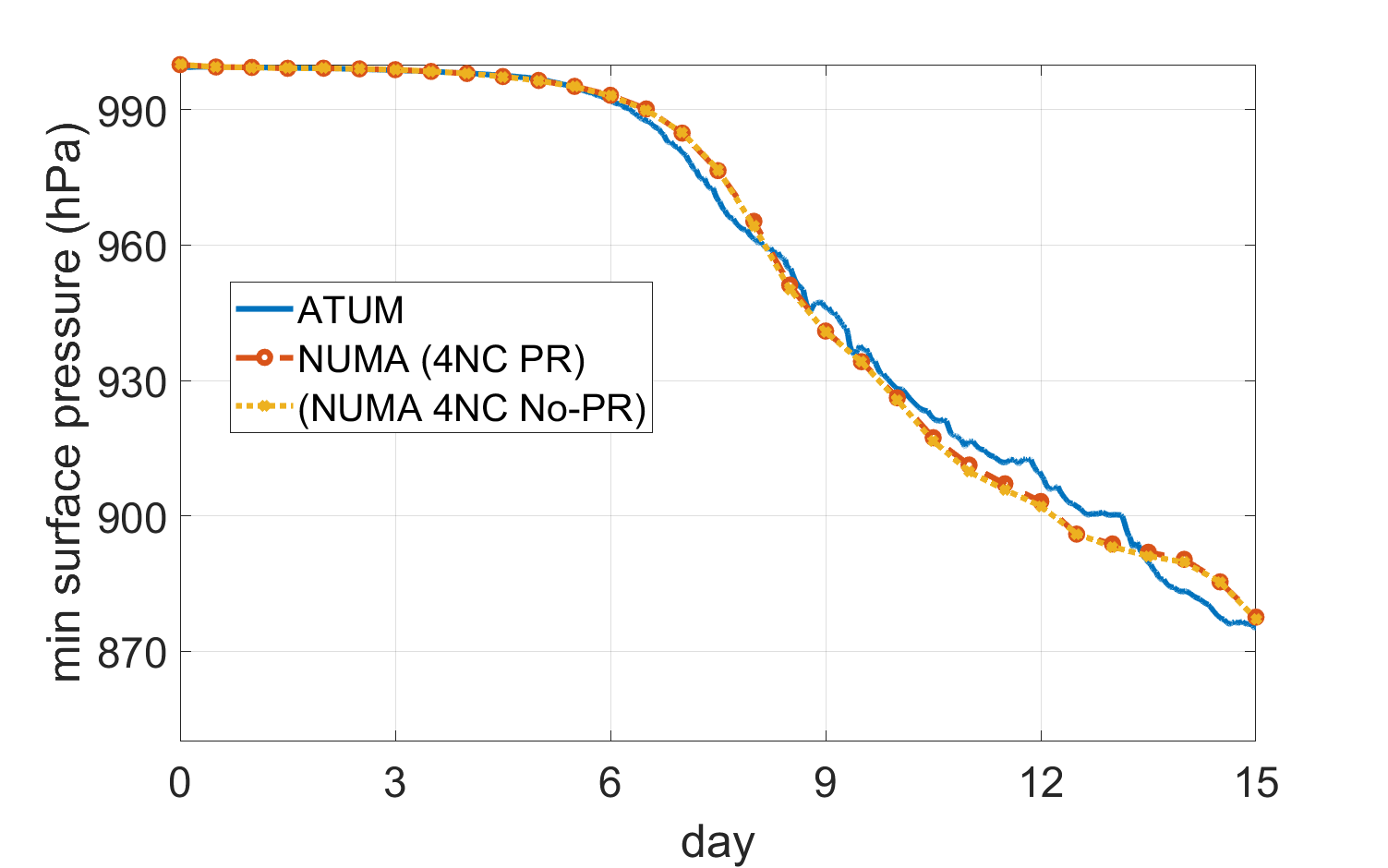}}
\subfigure[maximum horizontal velocities]{\includegraphics[width=3.0in]{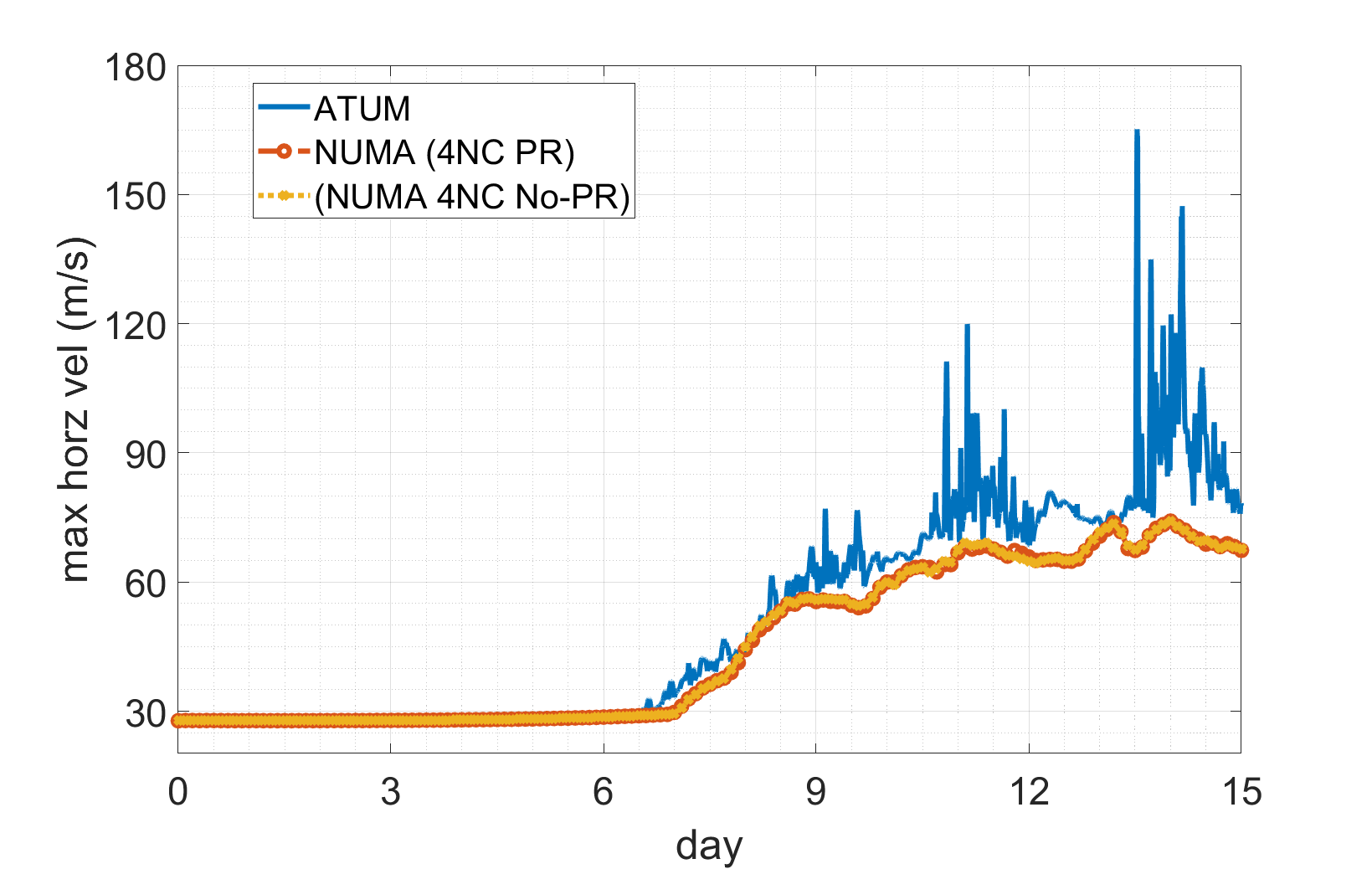}}
\caption{Minimum surface pressure (left) and maximum horizontal wind speeds (right) for the no-PR and PR forms in NUMA and for Atum.}
\label{fig:case600ts}
\end{figure}

Figure \ref{fig:case600ts} displays the minimum surface pressure (left panel) and the maximum horizontal wind speeds from the PR and no-PR NUMA DyCore runs and from the separate Atum run.  As noted in \cite{skamarock2021fully}, the minimum surface pressure diagnostic is relatively insensitive to discretization and the choice of parameters used in the stabilization scheme.  Hence, surface pressure diagnostics in Fig.~2a agree well among the PR and no-PR NUMA runs and Atum run, and also agree well with results from the seven nonhydrostatic and hydrostatic models shown in Fig.~3 in \cite{skamarock2021fully}.  
While the PR and no-PR NUMA runs produce similar maximum horizontal velocities in Fig.~\ref{fig:case600ts}b, the Atum run produces much larger horizontal velocities.  Comparing the results in Fig.~\ref{fig:case600ts}b to those in Fig.~4 of \cite{skamarock2021fully}, the NUMA horizontal winds are similar in magnitude to the reference MPAS simulation, while the Atum velocities behave like the low-hyperdiffusion ($1 \times 10^{13}$ m$^4$~s$^{-1}$) results presented in \cite{skamarock2021fully}.

\begin{figure}
\centering
\includegraphics[width=4in]{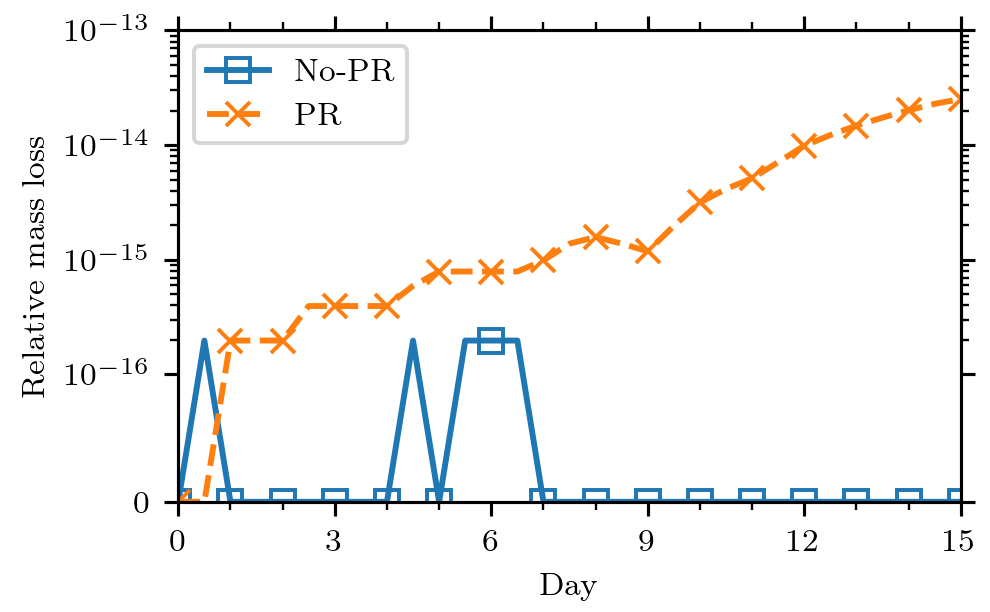}
\caption{Relative mass loss time series for the no-PR and PR forms of the DyCore for the Ullrich BI test case in NUMA.  The no-PR form conserves mass to machine precision.}
\label{fig:case600massloss}
\end{figure}

Figure \ref{fig:case600massloss} displays time series of domain-averaged relative mass loss for the no-PR and PR forms in NUMA.  The total mass at any given time $t$ is given by $M(t) = \int_{\Omega} \rho \, d\Omega$, and the relative mass loss is defined as
\begin{equation}
\frac{\delta M(t)}{M(0)} = \frac{| M(t) - M(0)|}{M(0)} .
\label{deltam}
\end{equation}
For details on the numerical evaluation of the total mass, see 
Sec. 3.3 in \cite{alves2023}.  The PR form \eqref{eq:set4ncpr} does not conserve mass and hence produces a secular mass loss during the course of the simulation.  In contrast, the no-PR form \eqref{eq:set4ncnopr} conserves mass to machine precision.  This mass conservation results from (a) using a conservative form of the continuity equation \eqref{eq:comassnopr}, (b) a weak imposition of the rigid (no-flux) BC, and (c) the
guarantee that the global integral of divergence vanishes using the SEM even under inexact numerical integration.

\begin{figure}
\centering
\includegraphics[width=6in]{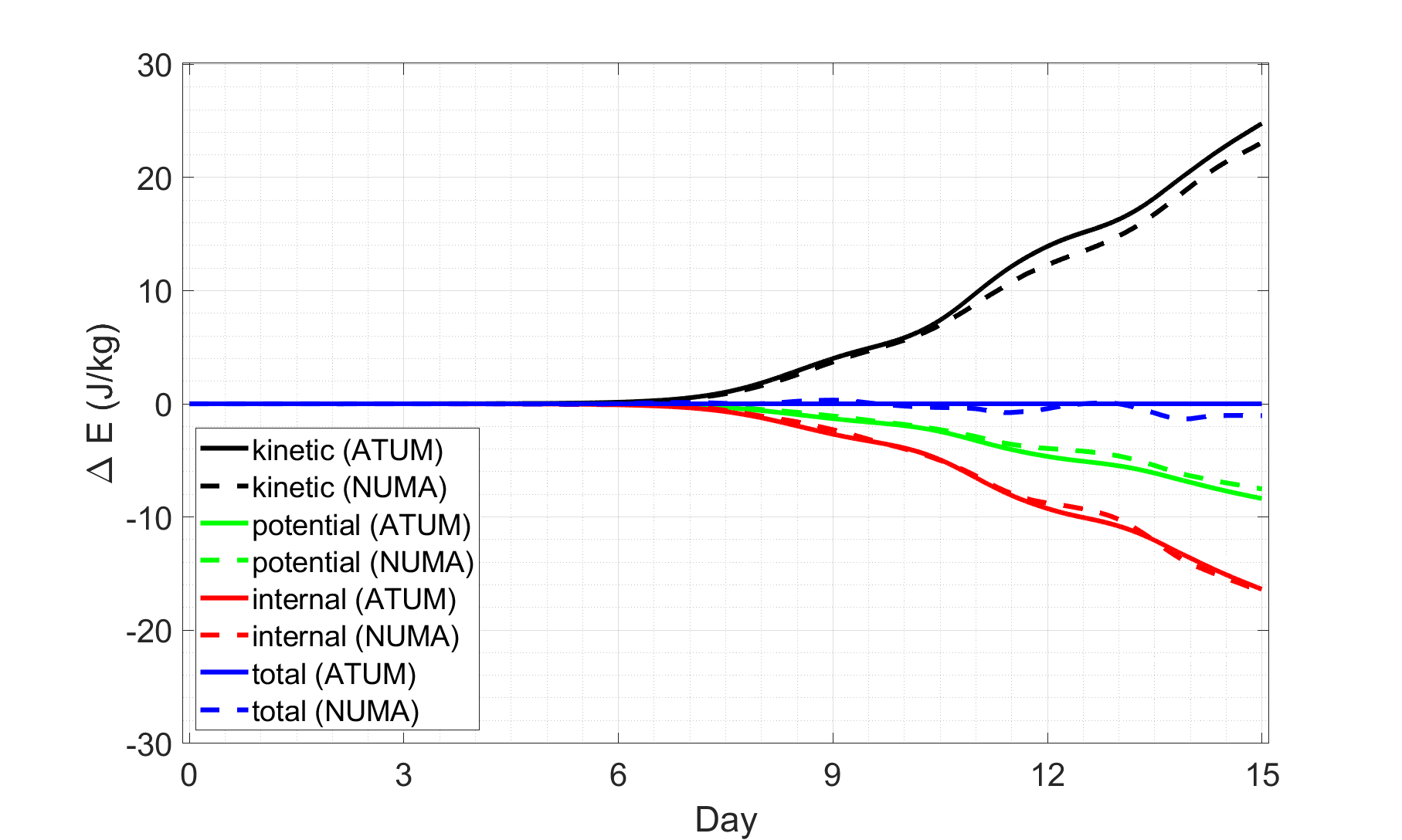}
\caption{The deviation in mean specific energy for the specific internal energy DyCore in NUMA and the total energy DyCore in Atum.  The deviation for the no-PR and PR forms in NUMA are similar, so only the no-PR energy budget is displayed.}
\label{fig:case600eb}
\end{figure}

Like many DyCores, newly SE-discretized forms of the equation sets \eqref{eq:set4ncpr} and \eqref{eq:set4ncnopr} in NUMA are not designed to conserve total energy.  As the BI dynamics evolve, internal and potential energies are transformed into kinetic energy (KE).  NUMA and many other dynamical cores numerically dissipate this KE via artificial dissipation (e.g., hyperdiffusion).   A useful diagnostic to study this energy transfer, proposed in \cite{skamarock2021fully}, is a time series of the deviations of kinetic, internal, and potential energies, where these quantities are as defined by (4)-(5) in Section 2b and domain integrated as for mass $M(t)$, such that $K(t)$, $I(t)$ and $P(t)$ are the domain-integrated forms of $E_k$, $E_i$ and $E_p$, respectively.
%
The mean specific values of these quantities are obtained by dividing by the total mass $M(t)$.  We then study absolute energy changes
from the initial value: e.g., $\Delta \overline{e}_k = K(t)/M(t) - K(0)/M(0)$.   Figure \ref{fig:case600eb} displays time series of these deviations in mean specific energy for the no-PR NUMA run.  Since the time series of NUMA PR results are almost identical to the no-PR results, those results are omitted for clarity.

From days 0--7, these energy deviations are all very small.  After the onset of the BI at day 7, kinetic energy is produced at the expense of the internal and potential energies.  Similar to Fig.\ 5 of \cite{skamarock2021fully}, there is a small decrease in total mean specific energy after day 8 since the dissipated kinetic energy is not accounted for in the total energy budget.  Since Atum does not dissipate kinetic energy, the total mean kinetic energy in the Atum run is larger than that in the NUMA runs.

The small loss in mean specific total energy due to the dissipation of kinetic energy by hyperviscosity in the NUMA simulation is comparable to the results reported in \cite{skamarock2021fully} after 15 days of simulation.  
Even though Atum conserves energy, it does so at the expense of simulation accuracy. The BI involves a downward cascade of energy that is interrupted by reaching the horizontal gridscale. Without hyperdiffusion, that energy aliases near the gridscale. 

\subsection{Hydrostatic Global Balanced State with Terrain}
\label{sec:labalancedflow}
Our second low altitude experiment involves initializing NUMA with a global atmospheric state in hydrostatic and gradient wind balance containing sinusoidal global-scale variation in terrain height (see \cite{jablonowski2006baroclinic} for details).  Here we zero out the BI perturbation term ($u_p=0$ in eq. (10) of \cite{jablonowski2006baroclinic}), such that the DyCore test maintains this balanced initial state over time.  Although this test case is formulated for shallow-atmosphere, hydrostatic dynamical cores in a pressure-based vertical coordinate, it is straightforward to initialize this balanced state in a terrain-following, height-based coordinate using Newton iteration.  The purpose of this test is two-fold: (1) to verify that the no-PR form of \eqref{eq:set4ncnopr} is stable and can conserve mass in the presence of topography and (2) to determine how the choice of metric terms influences mass conservation.  

This case was run with $u_0 =$~20~m~s$^{-1}$ in Eq. (2) of \cite{jablonowski2006baroclinic} using 6 elements per cube sphere panel and polynomial order 
$n=4$ with the ARK2 IMEX TI with $\Delta t$ = 100 s.  The upper BC is rigid for these tests and a terrain-following vertical coordinate is utilized with a mean vertical resolution of 1 km with a model top at 30 km.  Since no gravity waves are generated, a sponge BC was not necessary. 

\begin{figure}
\centering
\includegraphics[width=4in]{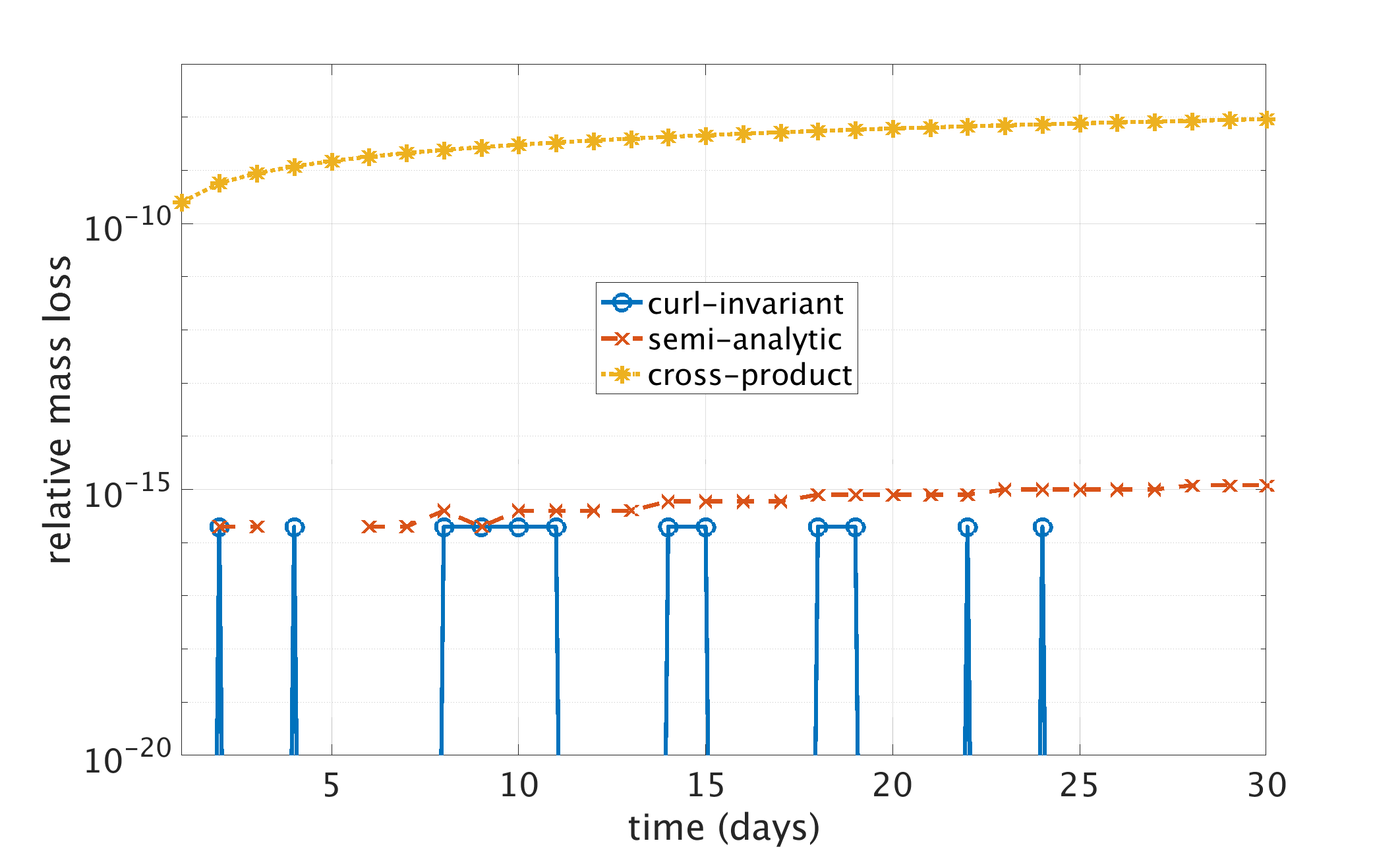}
\caption{A time-series of the relative mass loss over 30 days of integration for the hydrostatic Jablonowski-Williams BI test case using the
no-PR equation set using three choices of metric tensors: cross-product, semi-analytic, and curl-invariant. }
\label{fig:case6massloss}
\end{figure}

Figure \ref{fig:case6massloss} shows a 30-day time series of the relative mass loss \eqref{deltam} for NUMA runs using the no-PR DyCore with cross-product, semi-analytic and curl-invariant forms of the metrics, as described in Sections 3.2.1-3.2.3. The results show that CI metrics conserve mass up to machine precision using double precision floating point arithmetic, despite the presence of time-truncation error in the IMEX TI method. The simulation with the semi-analytic metrics produces a small mass-loss which increases with time. The cross-product metrics produce an unacceptably large mass loss that is five orders of magnitude larger than the corresponding simulation with the semi-analytic metrics.  We found that the gradient of the geopotential in \eqref{eq:momnopr} needed to be computed \emph{numerically} using the mimetic SEM to achieve mass conservation using the CI metrics.

These results indicate that the no-PR form \eqref{eq:set4ncnopr}, when discretized with a mimetic SEM utilizing the CI metrics specified 
by \eqref{eq:curl-invariant-metrics}, can conserve mass with or without topography.  We repeated this experiment with the ARS343 IMEX TI \cite{gardner2018implicit}, which produced similar results.  Since CI metrics do not have any more computational overhead than the cross-product or semi-analytic metrics, these results indicate that the CI metrics are the preferred metrics for global-scale simulations.
\section{High-Altitude Numerical Experiments}
\label{sec:haresults}
We implemented the PR form of the DyCore equations \eqref{eq:set4ncpr} in NEPTUNE and ran a series of idealized high-altitude tests.  The first test, described in Section \ref{sec:balancedflow} involves a steady-state balanced zonal flow based on specified vertical profiles of $T$, $R$ and $\gamma$ on a nonrotating Earth.  We initialize NEPTUNE with this balanced atmospheric state to test the ability of the new DyCore to maintain this state over time in the presence of varying $R$ and $\gamma$ due to variable upper atmospheric composition. In section \ref{sec:schar}, we add terrain to these experiments via an isolated idealized obstacle located at the equator. The peak obstacle height is kept small so that gravity waves forced by flow over the obstacle remain linear up to high altitudes, facilitating comparisons with linear solutions over a deep atmospheric vertical domain. Both constant and variable $R$-$\gamma$ states are used to assess impacts on simulated wavefields.


%
\subsection{Balanced Zonal Flow}
\label{sec:balancedflow}
%

An analytic steady-state solution in exact hydrostatic and gradient-wind balance on the sphere can be derived using the compatibility relations presented in \cite{white2008generalized} for a non-rotating planet
without terrain.  That solution consists of a balanced zonal wind of the form
\begin{equation}
u (z, \phi) = u_{eq} \sqrt{\frac{R(z) T(z)}{R_0 T_0}} \left(1 + \frac{z}{a} \right) \cos \phi ,
\label{eq:zonalwind}
\end{equation}
where $u_{eq}$ is the surface zonal wind velocity at the equator, 
$T_0 = T \left(z_s \right)$, $R_0 = R\left(z_s \right)$, $z_s$ is the surface height, and $a$ is the radius of the earth.  The corresponding hydrostatically balanced pressure is 
\begin{equation}
p(z,\phi) = p_s \exp \left( \frac{u^2_{eq}}{R_0 T_0} F_2 (z) \cos^2 \phi - \frac{u^2_{eq}}{2 R_0 T_0} \sin^2 \phi -  F_1(z) \right) ,
\label{eq:pressbal}
\end{equation}
where $p_s$ is the surface pressure, $F_2(z) = z/a + 1/2 (z/a)^2$ and 
\begin{equation}
F_1 (z) = \int_{z_s}^z \frac{g(z')}{R(z') T(z')}  \, dz' .
\label{eq:f1}
\end{equation}
In the absence of terrain, the surface pressure $p_s$  is replaced by the equatorial surface pressure $p_0$ and the lower limit of integration in \eqref{eq:f1} is zero.
The hydrostatic integral \eqref{eq:f1} reduces to $F_1(z) = \Phi(z)/(R_0 T_0) = \hat{z}/H_{p}$ for an atmosphere of constant $T=T_0$ and gas constant $R_0$,
where $\hat{z} = \Phi(z)/g_0$ is the geopotential height and $H_{p} = R_0 T_0 /g_0$ is the pressure scale height.  The derivation of \eqref{eq:zonalwind}-\eqref{eq:f1} is independent of $\gamma$, and thus holds for any $\gamma(z)$ profile. 

\begin{figure}
\centering
\subfigure[Temperature]{\includegraphics[width=2in]{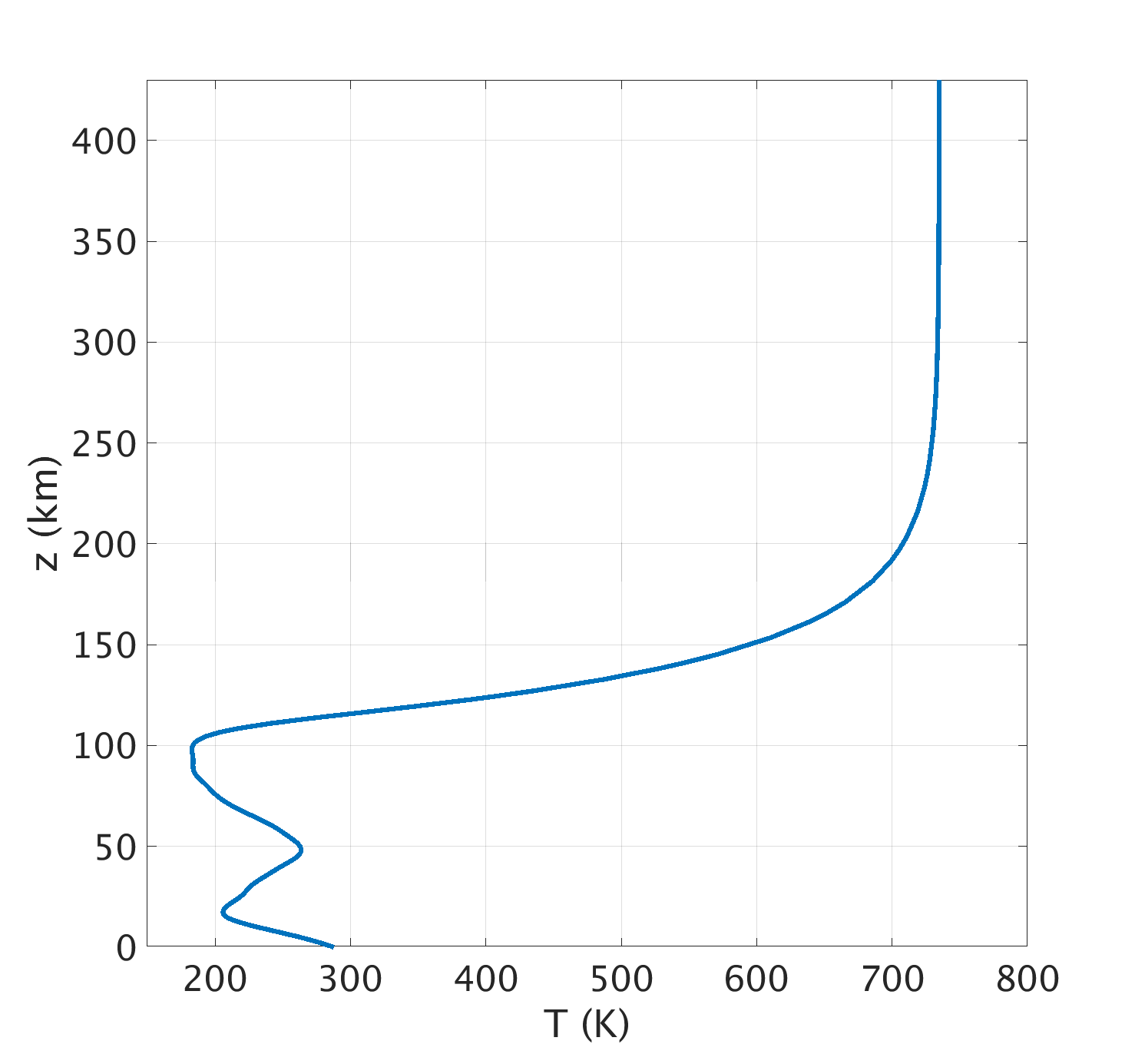}}
\subfigure[Specific Gas Constant]{\includegraphics[width=2in]{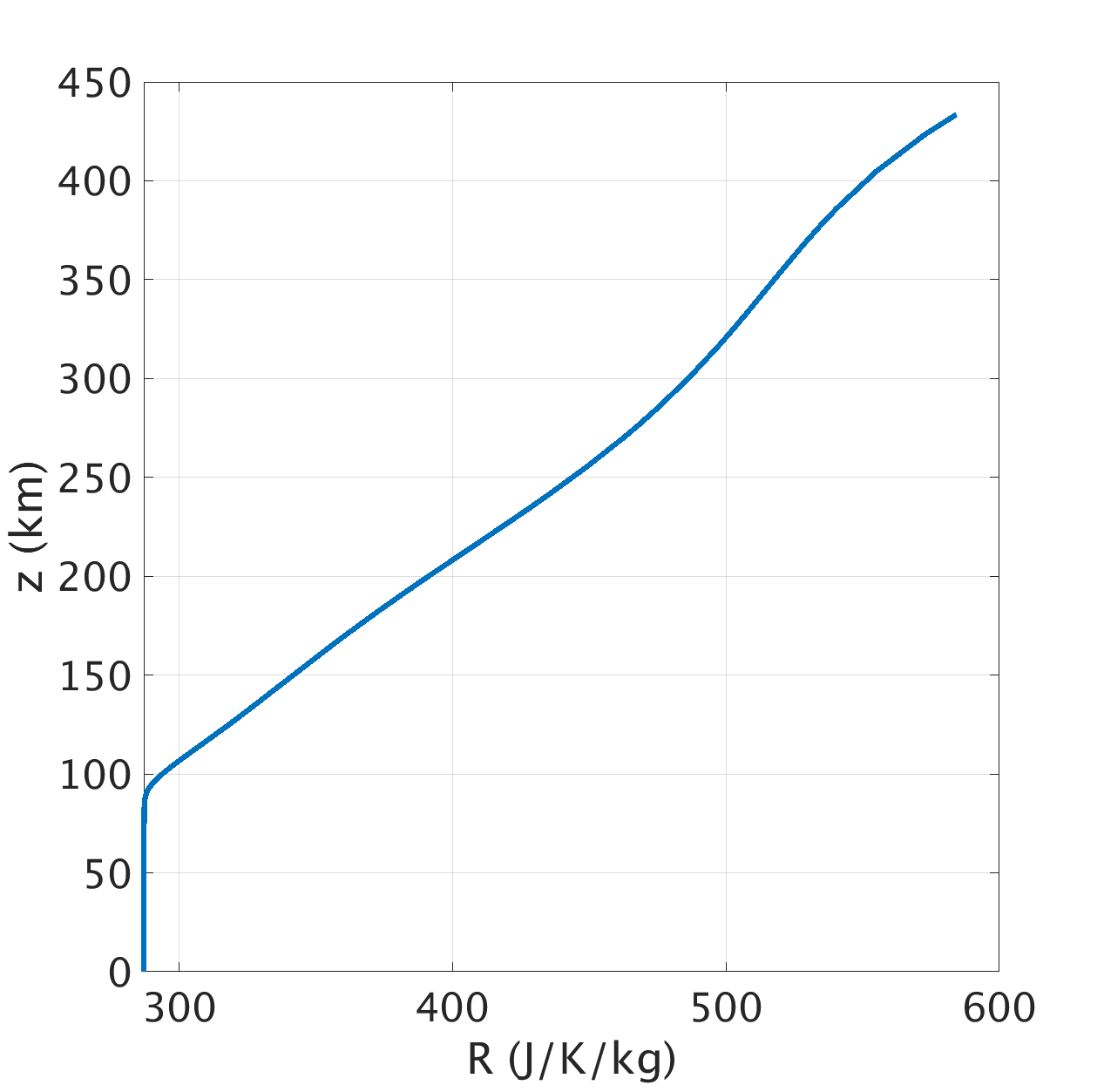}}
\subfigure[Ratio of Specific Heats]{\includegraphics[width=2in]{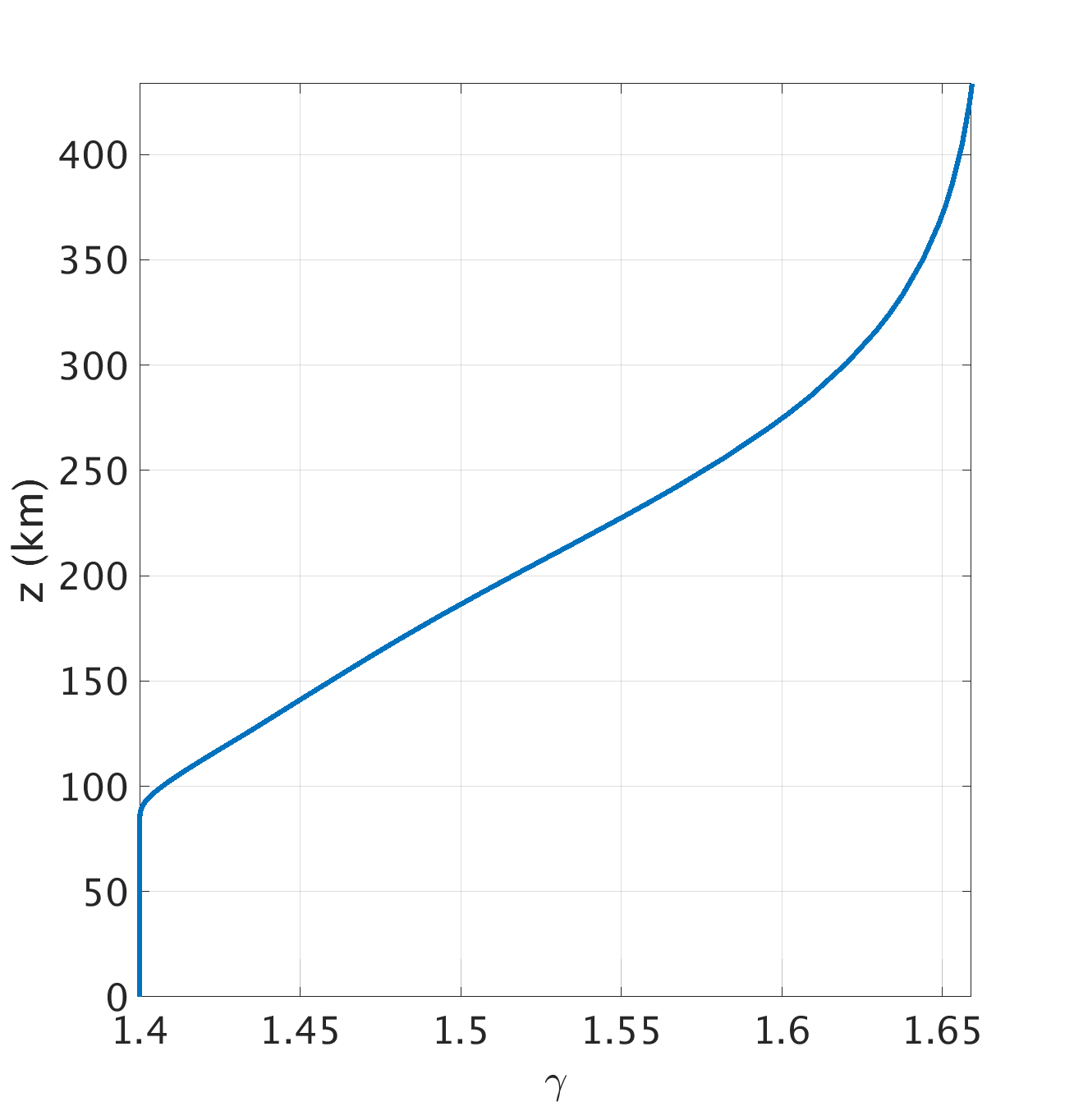}}
\caption{Vertical profiles of (a) temperature $T(z)$, (b) specific gas constant $R(z)$, and (c) ratio of specific heats $\gamma(z)$ used in the steady-state balanced zonal flow test case and subsequent orographic wave test case.}
\label{fig:case147profs}
\end{figure}

Figure \ref{fig:case147profs} displays the vertical profiles of $T(z)$, $R(z)$, and $\gamma(z)$ used in our numerical experiment with a model lid of 433 km.  Since these profiles are specified numerically, the hydrostatic integral \eqref{eq:f1} must be evaluated via a numerical quadrature scheme consistent with the SEM, as described in Appendix A.  The experiments were then run in NEPTUNE using $u_{eq} =$50~m~s$^{-1}$ in \eqref{eq:zonalwind} with no terrain and $p_0 =$1000~hPa in \eqref{eq:pressbal}. 
  
NEPTUNE was initialized with this steady state solution and ran using 20 SEs in the horizontal with $n=4$ order polynomials, which yields an 
approximate horizontal resolution of $\overline{\Delta x} \approx $ 125 km at the equator.  
A model top of 433 km was used with 46 elements and $n=4$ in the vertical, yielding a 185 level stretched vertical grid.  The vertical grid, which has an average grid spacing of $\sim$ 5 km in the thermosphere, is designed to capture large-scale features, such as large scale heating and cooling, rather than gravity wave propagation.  A nonlinear HEVI TI method utilizing the ARS343 scheme \cite{gardner2018implicit} with a time step $\Delta t = $ 60 s was employed.  Anisotropic hyperviscosity using the scheme 
proposed in \cite{gubaetal2014} was applied to stabilize the SEM.  Rigid lower and upper BCs were applied.  

\begin{figure}
\centering
\includegraphics[width=4in]{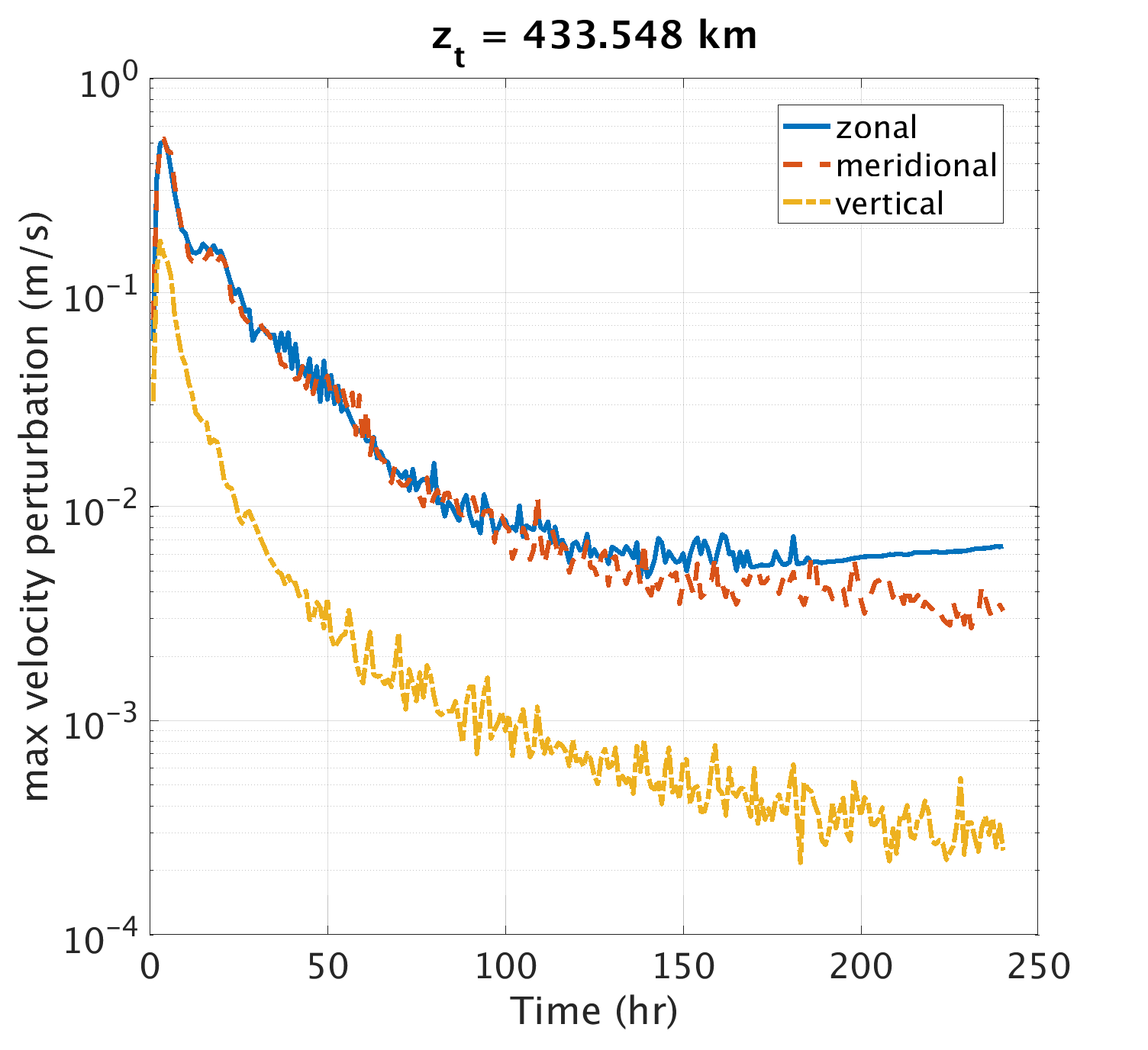}
\caption{Time series of maximum velocity perturbations for the balanced zonal flow test case specified by \eqref{eq:zonalwind} and \eqref{eq:pressbal}.  Small discretization and aliasing errors accumulate during
the first several hours of the simulation, yielding error in the horizontal and vertical velocities.  These numerical errors are then dissipated by the anisotropic diffusion scheme.}
\label{fig:case147vels}
\end{figure}

This experiment was run out in NEPTUNE for 10 days (240 h).  Figure \ref{fig:case147vels} plots the 10-day time series of maximum zonal, meridional and, vertical velocity derivations from the $t=0$ state. Small discretization and aliasing errors accumulate during
the first several hours of the simulation, then reduce substantially with time over the remainder of the simulation.  The largest velocity perturbations occur near the upper boundary, where small densities magnify amplitudes of any wave-related noise generated by small imbalances lower down.  These numerical errors are then dissipated by the anisotropic hyperdiffusion scheme, producing a small, residual error at the end of the simulation.  

Subsequent numerical experimentation was performed to reduce the discretization error in Figure \ref{fig:case147vels}.  Although this
experiment ran stably using a wide range of values for the hyperdiffusion coefficients, choosing too
small of a parameter produced unacceptably large velocity perturbations.  

It proved necessary to include surface integral (flux) terms in the hyperdiffusion scheme at the lower and upper boundaries.  These boundary terms, which result from 
performing integration by parts on the Laplacian operator, are proportional to the normal component of the gradient of each prognostic variable.  Since the zonal velocity in \eqref{eq:zonalwind} has linear shear in the vertical, this term must be accounted for to represent hyperdiffusion properly at the physical grid boundaries.  



\subsection{Orographic Gravity-Wave Test Case}
\label{sec:schar}
%
Our next test case used the balanced zonal flow from the previous section as the background state for an orographic gravity-wave simulation.  A balanced atmosphere with vertically varying temperature, specific gas constant, and ratio of specific heats was initialized using the profiles displayed in Figure \ref{fig:case147profs}.
Idealized terrain was introduced at the equator having the analytical form
\begin{equation}
h(\lambda,\phi) = h_0 \exp\left[ \frac{-d^2(\lambda,\phi)}{d_0^2} \right], 
\label{eq:hschar}
\end{equation}
where the great-circle distance from the mountain peak, centered at 
$(\lambda_c,\phi_c) = (\pi/4, 0)$, to any given point ($\lambda,\phi$) is
\begin{equation}
d(\lambda,\phi) = a \arccos \left[ \sin \phi_c \sin \phi +
               \cos \phi_c \cos \phi \cos \left( \lambda - \lambda_c 
               \right) \right] . 
							\label{eq:raxi}
\end{equation}

We choose a peak terrain height of $h_0 =$~0.2~m and a width $d_0 =$~500~km.  The small $h_0$ is chosen to keep steady-state wave solutions linear over the 0-300~km vertical domain.  The surface pressure $p_s$ was computed using the MSIS model \citep{emmertetal2020}.  To prevent reflection of gravity waves at the upper boundary, an implicit sponge of the form described in \cite{klempetal2008} is utilized.  The sponge depth is 100 km and the 
peak-damping time-scale is 0.25 s.  In order to suppress initial transient waves, a transient sponge, similar to Eq. (26) in \cite{klemp2021}, is applied at all altitudes with time-dependent coefficients.  As the simulation progresses, the sponge smoothly ramps down with respect to time as
\begin{equation}
\beta(z,t) = \beta_g (z) + \exp(-t/t_s) \beta_t (z),
\label{eq:bp}
\end{equation}
where $\beta(z,t)$ is the Rayleigh damping coefficient (units of $\mbox{s}^{-1}$) applied to the vertical momentum equation, $\beta_g(z)$ is an implicit gravity-wave-absorbing upper-level sponge with no time-dependence, 
\begin{equation}
\beta_t(z) = \beta_{tmax}^{-1} \sin^2 \left(\frac{\pi z}{2z_t} \right)
\label{eq:bptr}
\end{equation}
is an additional transient component that acts over the entire domain and targets acoustic waves, and $t_s = 5400$ s is the e-folding time scale that governs the ramp-down of the transient sponge.  The peak time-scale is $\beta_{tmax} = $ 0.25 s.  A Richardson number-based hyperdiffusion scheme was applied to suppress any destabilizing vertical grid-scale noise.

This experiment was performed in NEPTUNE with 24 elements per side of each cubed-sphere panel.
Vertical resolution of $\sim$ 1.1 km
was used using 79 vertical spectral elements with order $n=4$ order polynomials.  To ensure that all the gravity waves had sufficient time to propagate into the upper atmosphere to establish a steady-state solution, the experiment was integrated  to 240 h (10 days).  See Appendix B for verification of a steady-state solution after 240~h.

\begin{figure}
\centering
\includegraphics[width=3.0in]{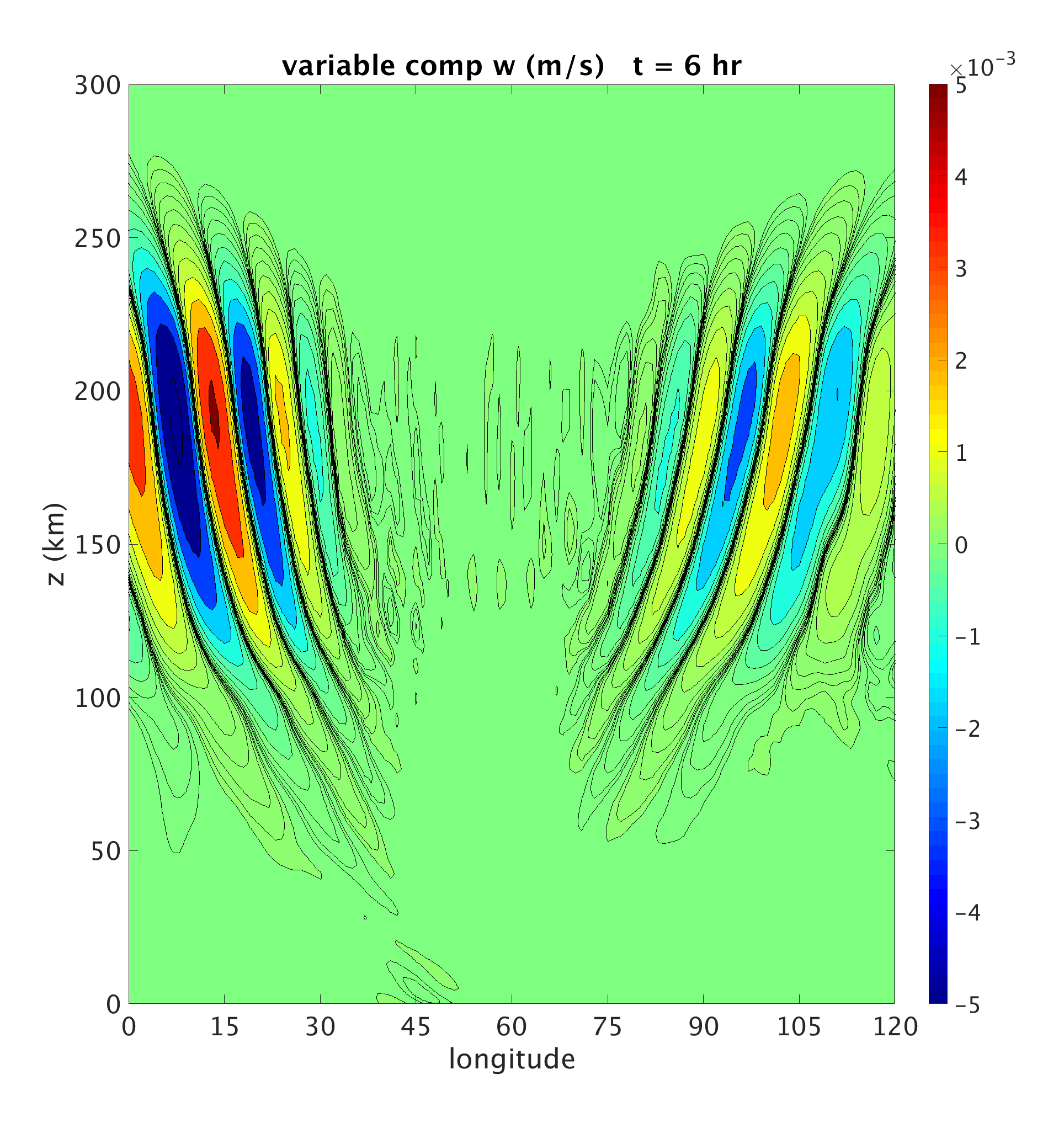}
\includegraphics[width=3.0in]{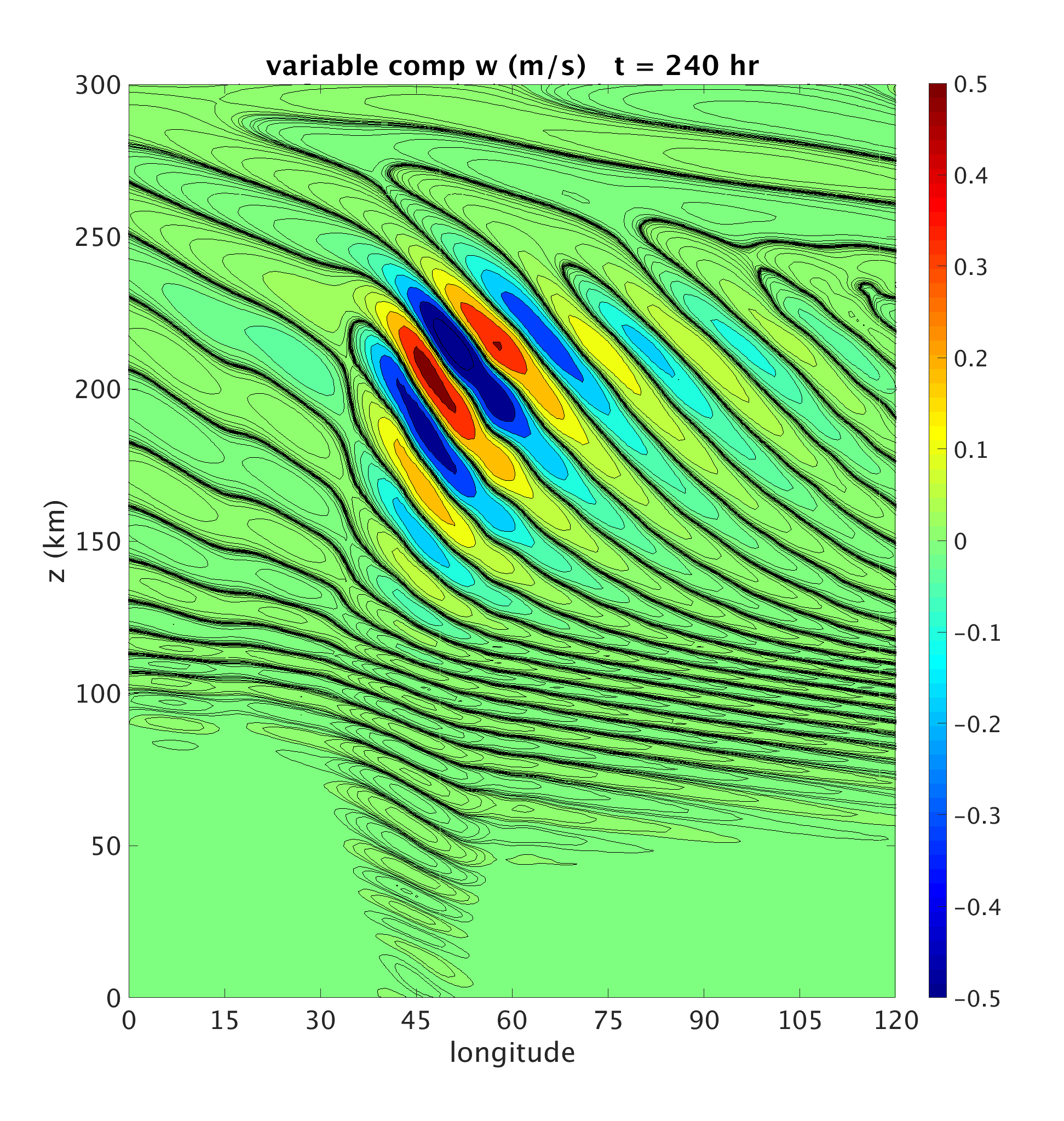}
\caption{Longitude-altitude cross sections along the equator of NEPTUNE vertical velocity $w$ (see color bars, units m~s$^{-1}$) for the orographic gravity-wave test case
with variable composition at (a) 6 h and (b) 240 h.  In order to display the wide dynamic range from the ground to the thermosphere, the $w$ contours are logarithmically-spaced from $10^{-5}$ m~s$^{-1}$ to 0.5 m~s$^{-1}$.  Panel (a) is dominated by acoustic waves, while panel (b) shows a steady-state gravity wave response.} 
\label{fig:varcompw}
\end{figure}

Figure \ref{fig:varcompw} displays longitude-altitude cross sections of vertical velocity $w$ at the equator at (a) 6 h (Fig.~\ref{fig:varcompw}a) and (b) 240 h (Fig.~\ref{fig:varcompw}b).  The response at 6 h is dominated by transient acoustic waves due to slight imbalances in the large-scale state with imposed terrain.  While acoustic transients are present in low-altitude DyCore tests involving terrain \cite{ullrich2017dcmip2016}, over the high-altitude domain considered here they attain large amplitudes in propagating rapidly to upper model levels, as shown in Fig.~\ref{fig:varcompw}a.  The transient sponge \eqref{eq:bp} is necessary to damp these high-amplitude waves, which would otherwise destabilize the simulation.  Similar behavior has been observed in other nonhydrostatic high-altitude models \cite{klemp2021}.  

In contrast, vertical velocities $w(\lambda,\phi,z,t)$ in Fig.~\ref{fig:varcompw}b 
at $\phi=0$ (equator) and $t =$~240~h are dominated by a steady-state 
gravity-wave response due to a persistent equatorial surface flow of $u_{eq} =$~50~m~s$^{-1}$ 
across the obstacle. Despite the small peak obstacle height of $h_0 =$~0.2~m, 
wave amplitudes grow secularly with height due to decreases in $\rho$ with height
to become large at upper levels. Amplitudes
approach 1~m~s$^{-1}$ at $z \sim$~200~km before decreasing in amplitude above that level
due to damping within the sponge layer. Downstream
of the obstacle peak at $\lambda_c =$45$^{\circ}$ there is significant wave activity, 
indicating nonhydrostatic modifications to the wave dynamics. 

\begin{figure}
\centering
\includegraphics[width=3.0in]{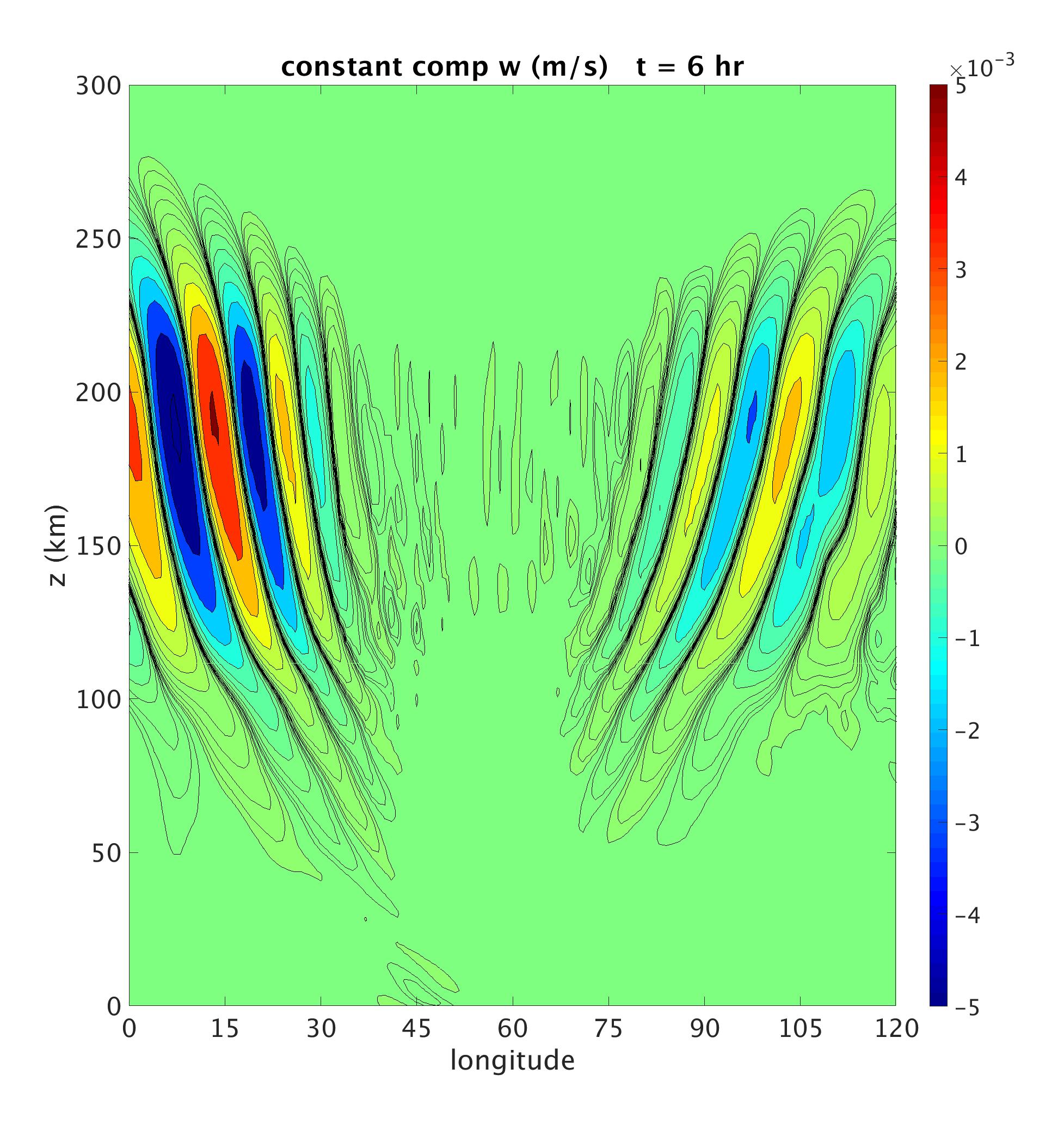}
\includegraphics[width=3.0in]{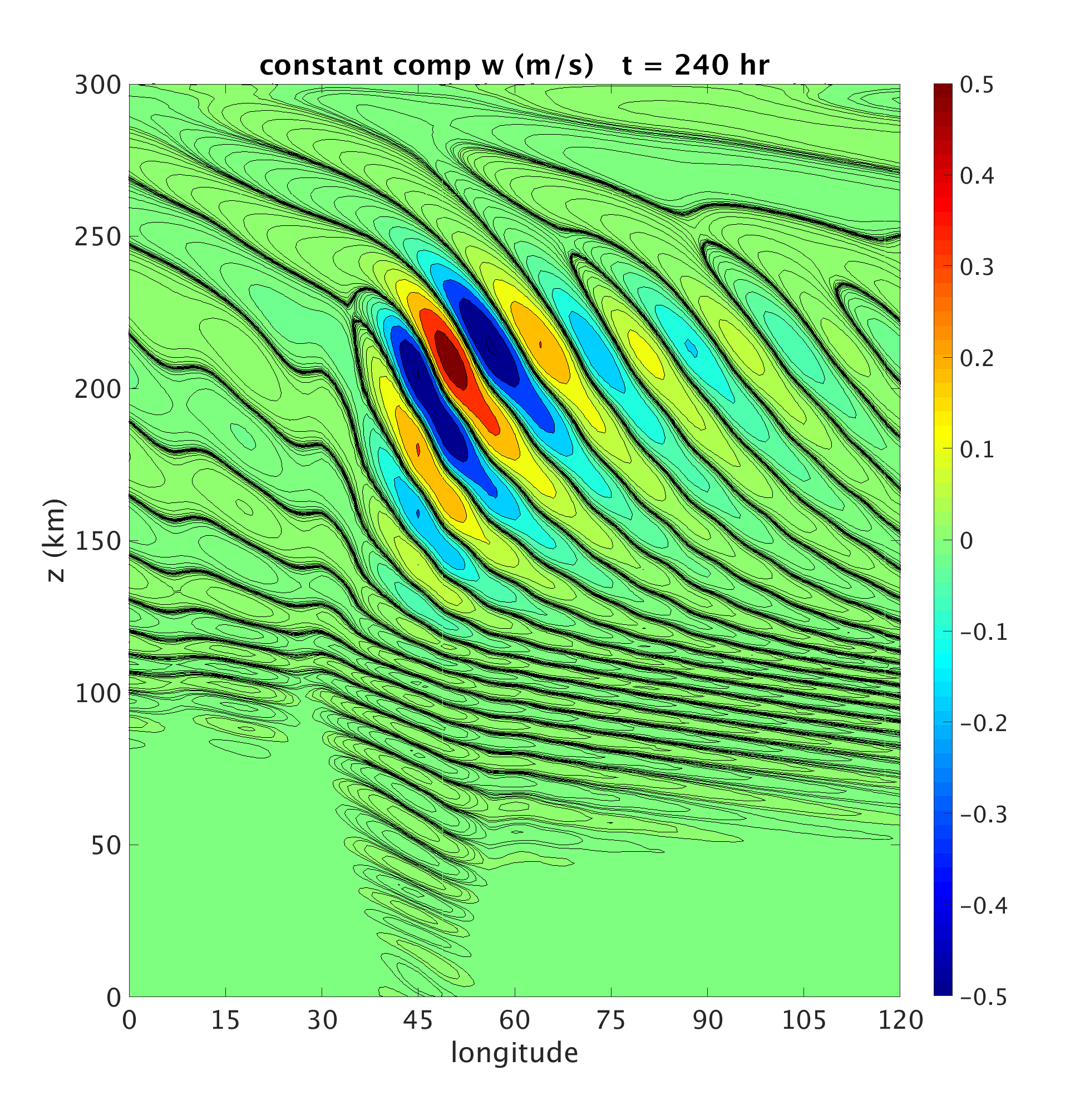}
\caption{Longitude-altitude cross sections along the equator of NEPTUNE vertical velocity $w$ (see color bars, units m~s$^{-1}$) for the orographic gravity-wave test case
with constant (uniform) composition at (a) 6 h and (b) 240 h.  Panel (a) is dominated by acoustic waves, while panel (b) shows a steady-state gravity wave response.} 
\label{fig:uniformcompw}
\end{figure}

\begin{figure}[tbh]
\begin{centering}
\includegraphics[width=3in]{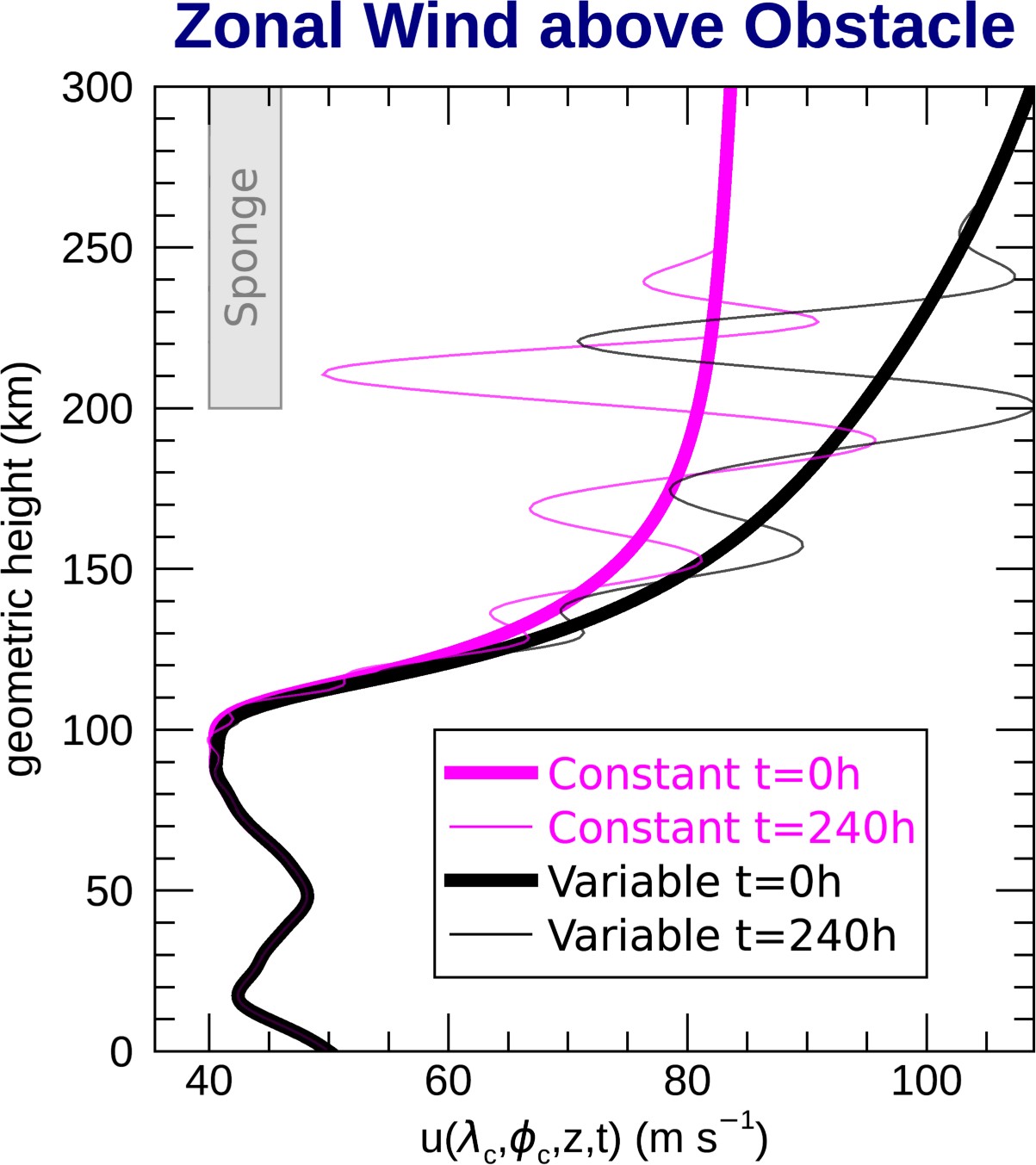}
\caption{Profiles of NEPTUNE zonal wind $u(\lambda_c,\phi_c,z,t)$ directly over the obstacle 
peak at $(\lambda_c,\phi_c)$ for a background state with 
variable (black) and constant (pink) $R$ and $\gamma$. Thick curves show winds at
$t =$~0~h given by the balanced solution (26). Thin curves show corresponding profiles at
$t =$~240~h showing large-amplitude gravity-wave oscillations superimposed upon these
background wind profiles.} 
\label{fig:uwind}
\end{centering}
\end{figure}

To assess the relative impacts of variable $R$ and $\gamma$ on these upper-level wavefields,
we ran a companion experiment that replaced the $R(z)$ and $\gamma(z)$ 
profiles in Figs. 7a and 7b, respectively, with profiles that applied the constant low-level 
values of $R = $~287.04~J~kg$^{-1}$~K$^{-1}$ and $\gamma = $1.4 throughout the domain.
Figure \ref{fig:uniformcompw} displays longitude-altitude cross sections of vertical velocity $w$ at the equator at 
(a) 6 h (Fig.~\ref{fig:uniformcompw}a) and (b) 240 h (Fig.~\ref{fig:uniformcompw}b). 
Although all other parameters were unchanged, Fig.~\ref{fig:uwind} shows that these changes 
in $R$ produced a large change in the balanced zonal wind profile above the mountain 
via (26), and hence to
the gravity-wave oscillations that become superimposed upon those background winds after 240~h.
These changes are evident in the corresponding cross-sections of 
vertical velocity $w(\lambda,0,z,t)$ at $t = $6~h (Fig.~10a) and $t =$~240~h (Fig.~10b).
When compared to the corresponding 
variable $R$ and $\gamma$ solution in Fig.~9b, the gravity waves in Fig.~10b have
noticeable differences at upper levels. For example, 
the weaker background winds for constant $R$ and $\gamma$ yield a smaller vertical 
wavelength at upper levels, consistent with the expected approximate linear dependence 
of vertical wavelengths on local $u$ through the gravity-wave dispersion relation.

To assess and validate these NEPTUNE gravity-wave results more
quantitatively, we derived corresponding time-dependent three-dimensional 
linear gravity-wave solutions from 
a Fourier-ray (FR) code, described in Appendix~B. FR solutions were derived
on a 1024$\times$1024 Cartesian $(x,y)$ domain with 40~km horizontal grid spacing 
and on 250 vertical levels using uniform 1~km vertical grid spacing. The obstacle (29) 
was positioned at the center of the $y$ domain but towards the left (westward) side of 
the $x$ domain to allow ship-wave patterns to evolve downstream without spurious 
upstream wraparound. A small amount of vertical diffusive damping ($K_{zz} =$~0.1~m$^2$~s$^{-1}$)
was introduced to reduce wraparound contamination of the solutions \citep[for details, 
see Appendix of][]{eckermannetal2015}. 
Figure~11 displays longitude-altitude cross sections of the 240~h FR vertical velocity 
solutions for the variable and constant $R$ and $\gamma$ atmospheres, such that Fig.~11a
can be compared to Fig.~9b and Fig.~11b can be compared to Fig.~10b. 
Overall similarity of the wave solutions is evident.

To facilitate more quantitative comparisons, Figure~\ref{fig:wkbv} compares vertical
profiles of $\vert w(\lambda_c,\phi_c,z,t) \vert$ at $t =$~240~h from NEPTUNE and from the FR
solutions directly above the obstacle peak $(\lambda_c,\phi_c)$, 
allowing the phase, vertical wavelength scales and amplitudes to be 
compared simultaneously throughout the vertical domain.
Very close agreement is evident between the linear FR solution and the NEPTUNE results 
throughout
the undamped domain from $z =$~0--200~km, in terms of local phase, vertical wavelength
and amplitude of the vertical velocity field. There is a slight amplitude discrepancy
in the variable $R$ and $\gamma$ case at upper levels, which we believe orginates from
small errors due to simplifications inherent in the FR model. For example, the FR 
code does not fully incorporate spherical modifications to wave dynamics, does
not include the meridional variation in the background zonal wind (26), and
does not include evanescent tunneling of wave activity through turning points.
The results in Fig.~\ref{fig:wkbv} show that the new DyCore produces and maintains 
accurate gravity-wave solutions over a deep vertical domain for both a constant
and variable composition atmosphere in response to flow across a very slight surface 
obstacle of 0.2~m peak height, further verifying the fidelity of
these discetized deep-atmosphere DyCore dynamics.
\begin{figure}
\centering
\includegraphics[width=3.0in]{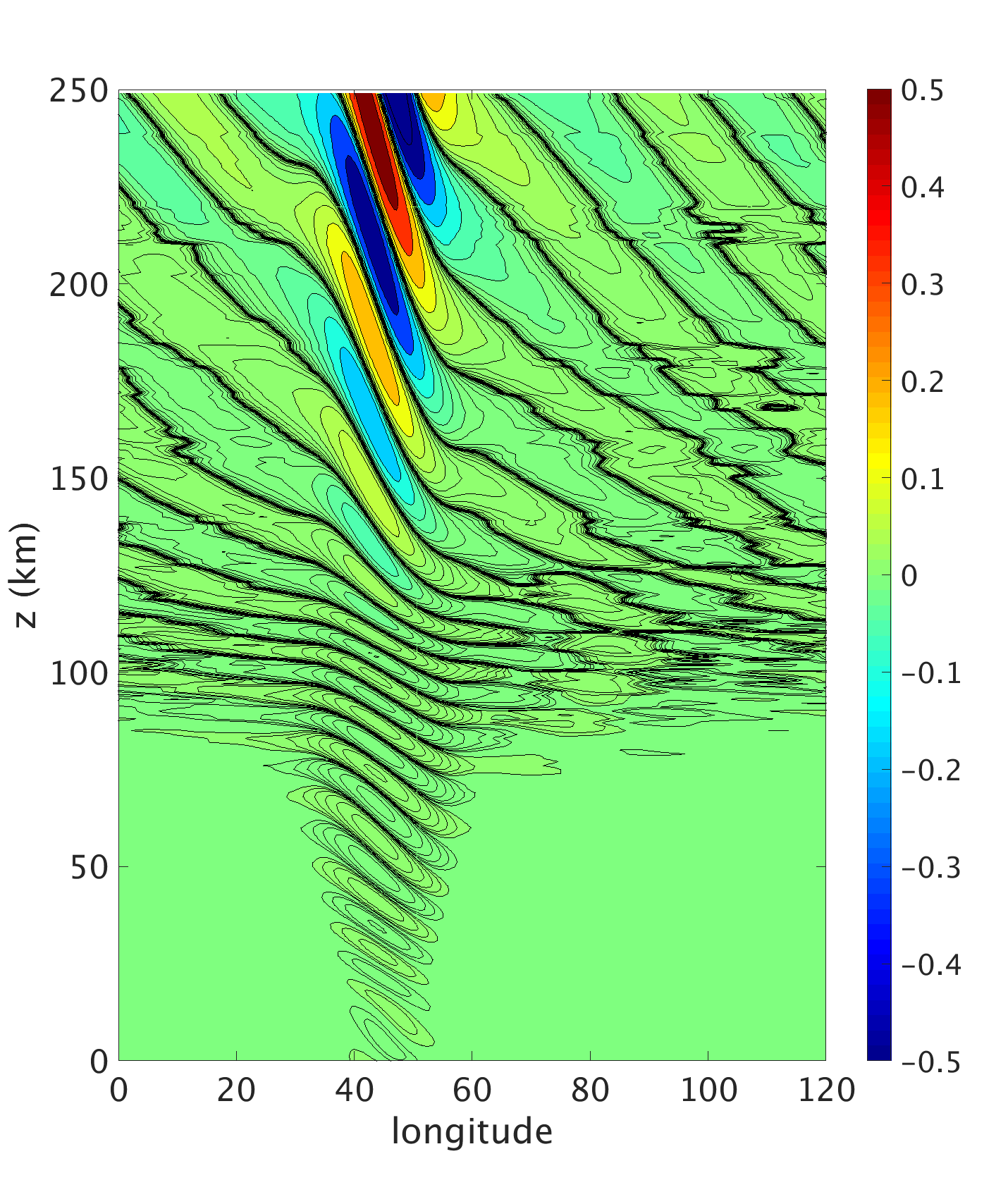}
\includegraphics[width=3.0in]{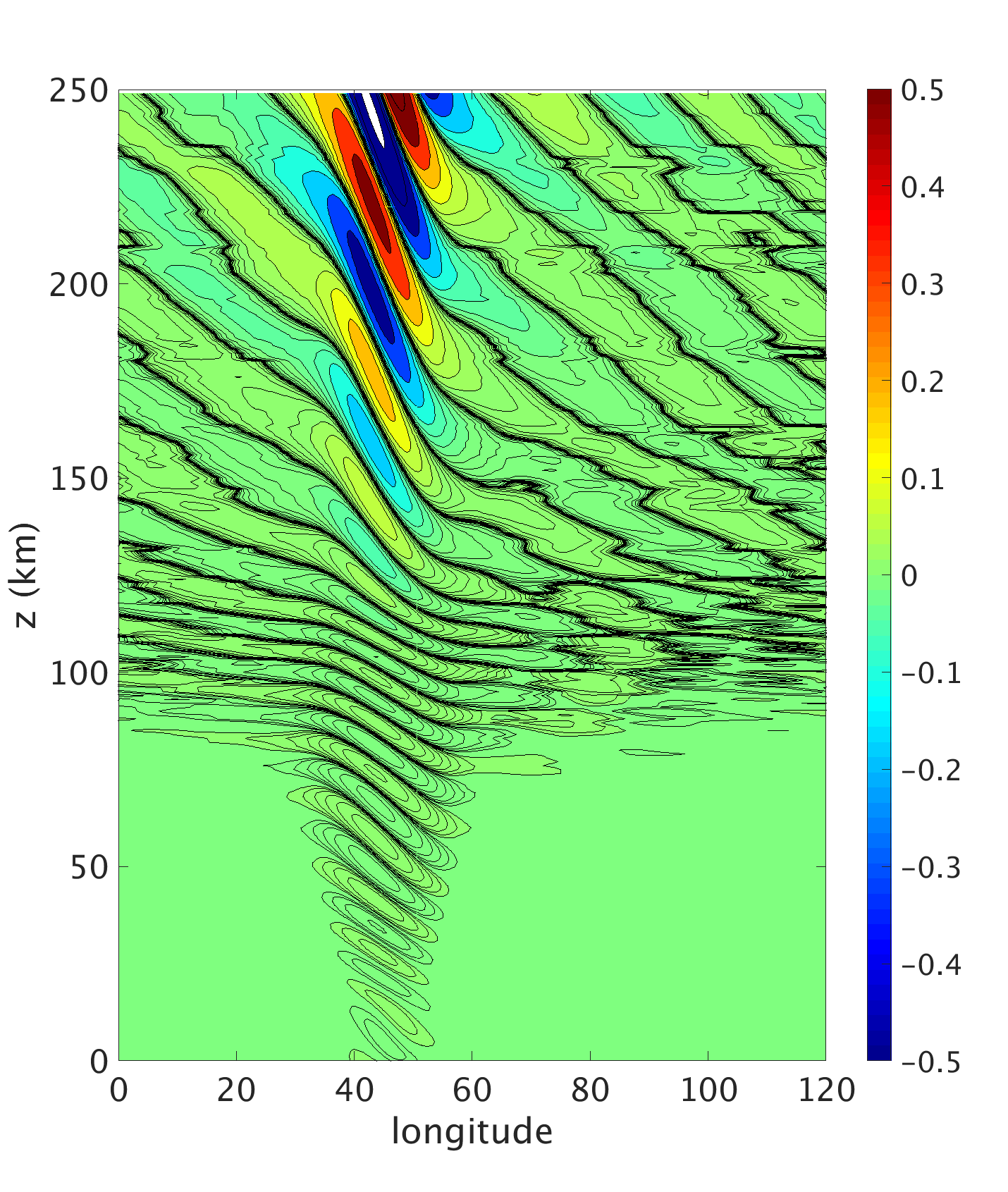}
\caption{Longitude-altitude cross sections along the equator of FR vertical velocity $w$ (see color bars, units m~s$^{-1}$) for the orographic gravity-wave test case with (a) variable composition and (b) constant composition at 240 h.} 
\label{fig:compfr}
\end{figure}
\begin{figure}[tbh]
\begin{centering}
\includegraphics[width=5in]{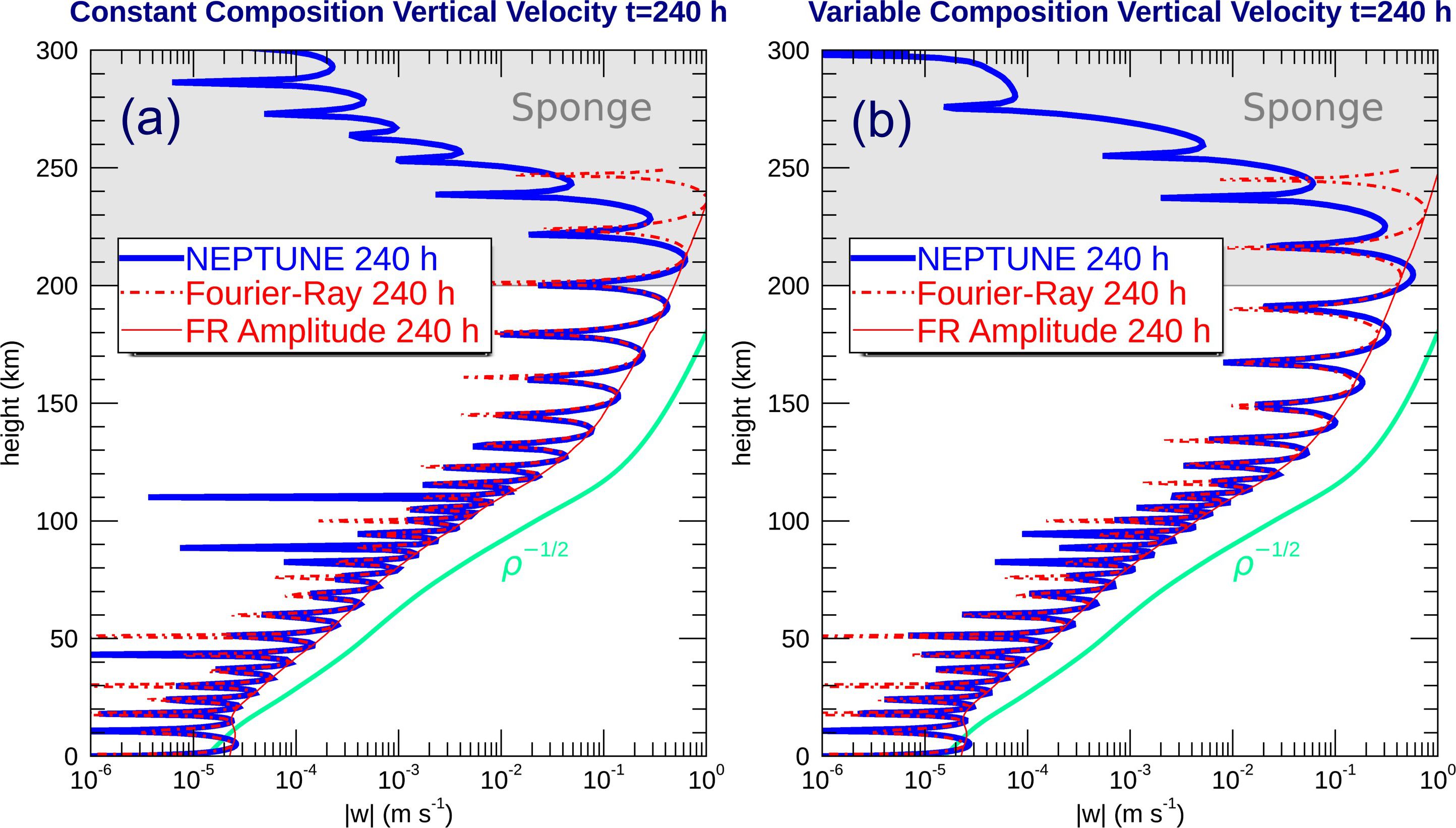}
\caption{Profiles of absolute vertical velocity 
$\vert w(\lambda_c,\phi_c,z,t) \vert$ directly above the obstacle peak at $(\lambda_c,\phi_c)$ and 
$t =$~240~h from NEPTUNE (blue solid curves) and the FR model (red broken curves) for (a) constant $R$
and $\gamma$ and (b) variable $R$ and $\gamma$. Red solid curves show the peak wave amplitude
of the FR solution (see Appendix B for details). 
Mint-colored curves show qualitative $\rho^{-1/2}$ vertical variation 
that leads to the secular increase in wave amplitude with height. 
NEPTUNE sponge layer is depicted in gray, which introduces dissipation to 
NEPTUNE solutions that causes them to diverge from FR solutions above 200~km.} 
\label{fig:wkbv}
\end{centering}
\end{figure}
\section{Energy Conservation}
\label{sec:energycons}
As shown in Sec.~\ref{sec:laresults}, the no-PR form \eqref{eq:set4ncnopr} conserves mass but not total energy when inexact integration is utilized.  In this section, we establish the condition under which total energy is conserved. The kinetic, potential, and internal energy per unit volume are given in \eqref{eq:etot2}.  If exact (spatial) integration and TI are used, then the product rule for the discrete gradient $\grad_d$ holds and we are justified in taking the time-derivative of the three components of energy \eqref{eq:etot2}, yielding
\begin{subequations}
\label{eq:energy_nopr/derivative}
\begin{equation}
\diff{E_k}{t}= \left( \frac{1}{2} \vc{u} \cdot \vc{u} \right) \diff{\rho}{t}  + \rho \vc{u} \cdot \diff{\vc{u}}{t} ,
\label{eq:energy_nopr/kinetic/derivative}
\end{equation}
\begin{equation}
\diff{E_p}{t}= \diff{\rho}{t} \Phi +  \rho \diff{\Phi}{t} ,
\label{eq:energy_nopr/potential/derivative}
\end{equation}
\begin{equation}
\diff{E_i}{t}= \diff{\rho}{t} e_i + \rho \diff{e_i}{t}.
\label{eq:energy_nopr/internal/derivative}
\end{equation}
\end{subequations}
Substituting the time-derivatives into \eqref{eq:set4ncnopr}, summing, and collecting terms yields
%
\begin{align}
\diff{\left(E_k + E_p + E_i \right) }{t} &= - \grad \cdot  \left[ \left( \frac{1}{2} \rho \vc{u} \cdot \vc{u} + \rho \Phi + \rho e_i  \right) \vc{u}  +  p \vc{u} \right] \nonumber \\
&= - \grad \cdot  \left[ \left( E_k + E_p + E_i + p \right) \vc{u}  \right] 
\label{eq:energy_conservation3}
\end{align}
which reduces to \eqref{eq:etot2}.  Hence, the no-PR form \eqref{eq:set4ncnopr} satisfies conservation of total energy provided that exact integration is utilized.  We note that the standard potential temperature equation does not share this property unless a certain vector-invariant form of the momentum equation is used.  Unlike the inexact integration adopted in this study, exact integration utilizes a non-diagonal mass matrix, and is hence more expensive.
\section{Discussion and Conclusion}
\label{sec:conclusion}
An important additional requirement is that a modified DyCore using \eqref{eq:ie2} should
be viable for operational tropospheric NWP as well,
particularly since thermospheric skill depends \textit{inter alia} on accurately
capturing the deep thermospheric propagation of multiscale wave disturbances generated
by tropospheric meteorology \citep{jacksonetal2019}. Note that \eqref{eq:ie2} is a
generalization of the temperature tendency equations already implemented in many
current operational NWP dynamical cores \cite[e.g.,][]{ritchieetal1995,untchhortal2004}.
Since these NWP temperature tendency equations approximate the exact $e_i$ equation \eqref{eq:ie2},
an $e_i$-based DyCore can potentially provide improved temperature prediction accuracy at
little additional computational overhead relative to existing DyCore temperature equations.
In addition to incorporating small
$c_v$ tendency terms due to composition changes due to water phase changes
in highly moist tropospheric environments, $e_i$ represents an exact
means for dealing with the associated latent heating terms within the DyCore
\citep[see, e.g., ][]{staniforthwhite2019,bowenthuburn2022a}. 

This paper has presented two ground-to-exosphere DyCores are presented for variable composition deep atmospheres, based on a specific internal energy equation that provides exact solutions to the first law of thermodynamics.  Product Rule (PR) and no Product Rule (no-PR) forms are discretized using the SEM and both an IMEX and HEVI time-integrator.  The DyCore is verified at low altitudes using two baroclinic instability (BI) test cases.  Deep atmosphere tests using NEPTUNE are presented for both a steady-state balanced zonal flow and an orographic gravity-wave test case.  A globally balanced atmospheric state from 0-300 km is accurately maintained in NEPTUNE  over 10 days of model integration. Introducing a small obstacle at the equator forces a steady-state orographic gravity-wave response from 0-200 km altitude that compares closely to linear numerical solutions for both constant and variable composition atmospheres.  When a no-PR form of the continuity equation is used with a consistent specific internal energy equation along with CI metrics and inexact integration, mass is conserved to machine precision out to 240~h with and without terrain.  The PR form does not conserve mass using inexact integration.
\section*{Acknowledgments}
Discussions with Jeremy Kozdon (NextSilicon), Sohail Reddy (Lawrence Livermore National Laboratory), and Simone Marras (Department of Mechanical and Industrial Engineering, New Jersey Institute of Technology) are acknowledged with pleasure. Maciej Waruszewski (Sandia National Laboratories) provided the Atum code used in this study. This work was supported by the Office of Naval Research Marine Meteorology and Space Weather program, and by the Space Environment Exploitation (SEE) program of Defense Sciences Office of the Defense Applied Research and Projects Agency (DARPA DSO). NEPTUNE and NUMA runs were facilitated by grants of computer time and resources at the Navy DoD Supercomputing Resource Center (DSRC) via a DoD High Performance Computing Modernization Program (HPCMP) Frontier Project.
\appendix  
\section{Recursive Inexact Integration}
\label{sec:indefsec}
To evaluate hydrostatic indefinite integrals like \eqref{eq:f1} numerically, we adopt the \emph{recursive inexact integration} approach used in the ClimateMachine 2.0 (CLIMA) model \cite{sridhar2021large} for maintaining hydrostatic balance.  Evaluation of \emph{definite integrals} within a CG framework using inexact integration is straightforward and efficient since quadrature points are collocated with interpolation points \cite[Chapter 4]{giraldo2020} here.  However, evaluation of indefinite integrals within an EBG framework is not straightforward.  

To illustrate why, let $f(x)$ be a continuous, real-valued function on the interval $[-1,1]$, or the reference interval for a spectral element.  Suppose we wish to evaluate the indefinite integral $F(x) = \int_{-1}^x f(z) \, dz$ at the set of Legendre-Gauss-Lobatto (LGL) points $x_i$, where $1 \leq i \leq N$, where $N = n + 1$, $n$ is the polynomial order, and $x_1 = -1$ and $x_N = 1$.  Letting $f_i = f\left(x_i \right)$, we approximate $f(x)$ in terms of Lagrange polynomials $h_i(x)$ with interpolation points $x_i$ and integration weights $\omega_i$.  Note that 
\begin{align*}
F \left( x_N \right) =& \int_{-1}^{1} f(z) \, dz \\
									   =& \sum_{i=1}^N f_i 	\int_{-1}^{1} h_i(z) \, dz ,\\
										 \approx& \sum_{i = 1}^N f_i \sum_{j = 1}^N \omega_j h_i \left(x_j \right), \\
										 \approx& \sum_{i = 1}^N f_i \omega_i,
\end{align*}
where we have used inexact integration in the third line and the cardinality property $h_i (x_j) = \delta_{i,j}$ in the fourth line, where $\delta_{i,j}$ is the Kronecker delta.  If we replace the definite integral with an indefinite integral in this calculation, inexact integration cannot be used to evaluate $\int_{-1}^x h_i (z) \, dz$.  Hence, we must construct a quadrature rule that is more accurate than inexact integration and hence appropriate for evaluation of \eqref{eq:f1}.

Consider the sub-interval $[x_i , x_{i + 1}]$ within the interval $[-1,1]$, where $1 \leq i \leq (N-1)$.  Let $R_i = (x_{i + 1} - x_i)/2$ be the radius of the sub-interval and let $m_i = (x_{i + 1} + x_i)/2$ be the mid-point. We then construct a sequence of integration points that covers $E_i$ via $x_{i,j} = R_i x_j + m_i$.  

An indefinite integral is decomposed into a series of definite integrals via:
\begin{equation*}
F(x_i) = \int_{-1}^{x_i} f(z) \, dz = \sum_{k=1}^{i-1} \int_{x_k}^{x_{k+1}} f(z) \, dz .
\end{equation*}      
These $(i-1)$ \emph{definite} integrals are evaluated via inexact integration
\begin{equation}
\int_{x_k}^{x_{k+1}} f(z) \, dz = R_k \sum_{j=1}^N f_{k,j} \omega_j ,
\label{subintegral}
\end{equation}
where 
\begin{equation}
f_{k,j} = f \left( x_{k,j} \right) = \sum_{m=1}^N f_m h_m \left( x_{k,j}  \right) .
\label{fkj}
\end{equation}
and $f_m = f \left(x_m \right)$ are function values at the original set of LGL points $x_m$.  Hence, we do \emph{not} require additional data to evaluate the indefinite integral.  However, note that there is no cardinality property for Lagrange polynomials in \eqref{fkj}, implying that \eqref{subintegral} requires evaluating a double sum. 
\section{Time-Dependent Numerical Fourier-Ray Solutions}
\label{sec:fr}
A linear time-dependent numerical solution to the orographic gravity-wave problem of
Sec.~\ref{sec:schar} is derived using a variant of the Fourier-ray (FR) method described in Sec.~
2b and the appendix of \cite{eckermannetal2015} and references therein. While idealized
vertical profiles of wind and temperature were used in that study, \cite{eckermannetal2016}
showed that the method produces accurate orographic gravity-wave solutions at
high altitudes when using realistic vertical profiles of background winds and temperatures
like those shown in Figs.~7 and \ref{fig:uwind}.

Wave properties are governed by compressible nonhydrostatic dispersion and polarization
relations.  Wave amplitudes are governed by a wave action equation that is
solved numerically in a mixed Fourier-height space. To include the effects of variable 
composition, the model's governing dispersion relation and wave action conservation equation were 
upgraded to use the variable-$\gamma$ forms derived by
\cite{eckermann2025}. Deep atmosphere effects are included by (a)
incorporating height-varying gravitational acceleration, (b) 
dilating horizontal wavelengths with height as $a/r$, where
$a$ is Earth radius, and (c) adding horizontal
geometrical spreading of wave amplitude due to a spherical Earth 
(see Appendix C of \cite{eckermann2025}).
Explicit shear and curvature terms were added to the dispersion relation
(see Eq.~(7) in \cite{broutmanetal2014}),
but produced imperceptible differences to wave solutions that omitted those terms.
Coriolis terms in the equations were deactivated consistent with the idealized
nonrotating NEPTUNE experiment.

\begin{figure}
\begin{centering}
\includegraphics[width=5in]{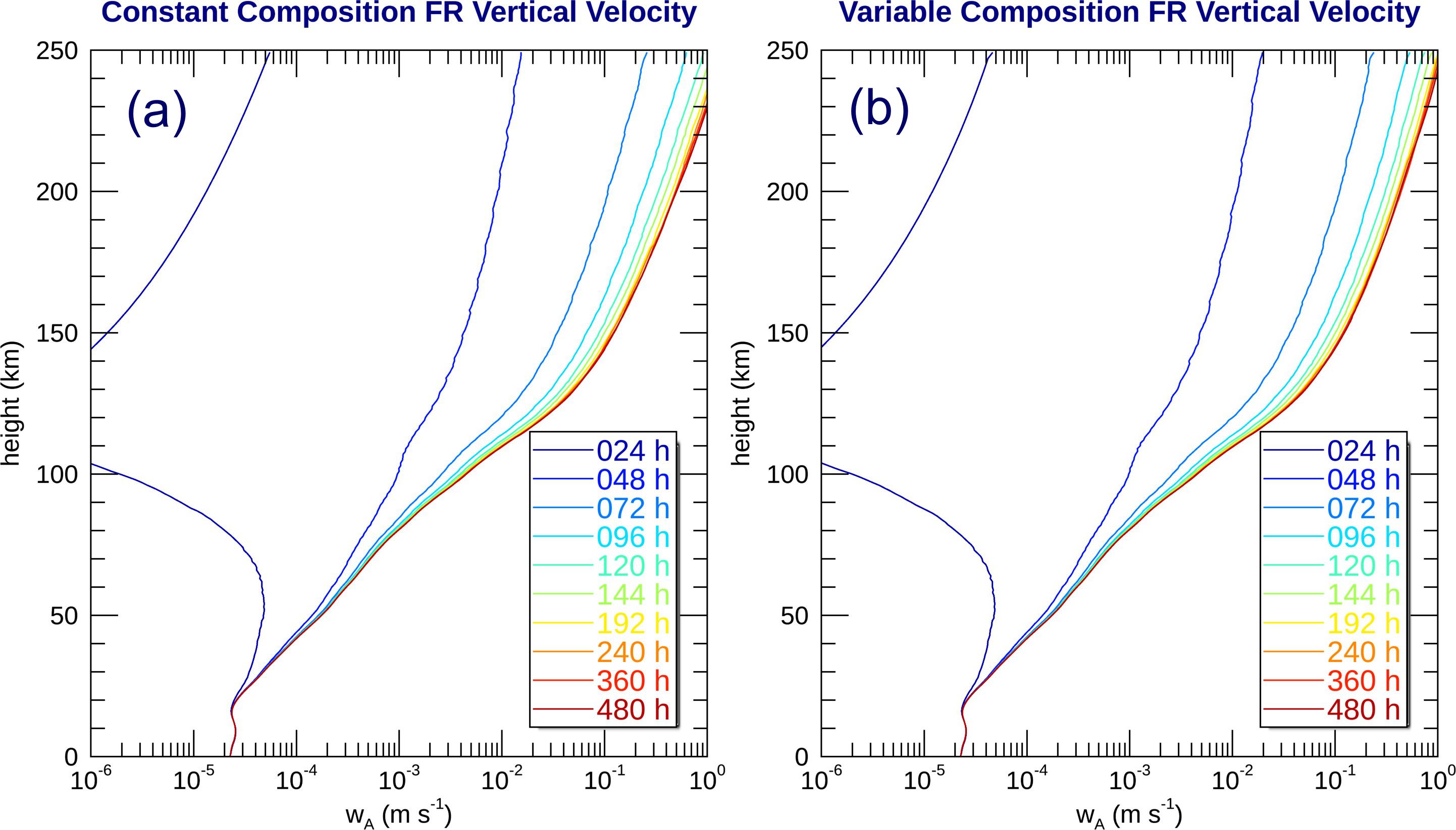}
\caption{Time evolution of $w_A$ above the obstacle peak from 0-480~h for FR solutions
derived using (a) constant and (b)
variable $R$ and $\gamma$ atmospheres.} 
\label{fig:wafr}
\end{centering}
\end{figure}

In addition to vertical velocity solutions
$w^\prime(x,y,z,t)$, peak wave amplitudes $w_A(x,y,z,t)$ were derived 
using a Hilbert transform technique described
in Sec.~2c of \cite{eckermannetal2015}. Figure~\ref{fig:wafr} shows the time evolution
of vertical profiles of $w_A(x_c,y_c,z,t)$ directly over the obstacle peak ($x_c,y_c$),
showing that a steady-state solution is approached throughout the vertical domain 
after 240~h. FR tests like these were used to tune the obstacle height $h_0$ to ensure that
steepness amplitudes $s_A$ did not exceed a linear criterion for 
wave breaking \cite{eckermannetal2015}, thereby ensuring linear and
stable NEPTUNE solutions.
\end{document}

%% file: FXG_Macros.tex
\usepackage[english]{babel} 

\usepackage{epsfig}

\usepackage{algorithmicx}
\usepackage{algpseudocode}

\usepackage{subfigure}
\usepackage{stmaryrd}
\usepackage{amsmath} 
\usepackage{color}
\usepackage{amssymb} 
\usepackage{amsfonts} 
\usepackage{mathrsfs}

\usepackage[T1]{fontenc}
\usepackage[utf8]{inputenc}
\usepackage[font=small,labelfont=bf,tableposition=top]{caption}
\usepackage{multicol}

\usepackage{float}
\usepackage{graphicx}
\input{epsf}

\newcommand{\comment}[1] {}

\newcommand{\diff}[2] {\frac{\partial #1}{\partial #2}}

\newcommand{\vc}[1]{\mbox{\boldmath$#1$\unboldmath}}

\newcommand{\be}{\begin{equation}}
\newcommand{\ee}{\end{equation}}
\newcommand{\bea}{\begin{eqnarray*}}
\newcommand{\eea}{\end{eqnarray*}}

